\documentclass[11pt]{article}
\usepackage[utf8]{inputenc}
\usepackage{authblk}
\usepackage{amsmath}
\usepackage{graphicx}
\usepackage{setspace}
\usepackage{xcolor}
\usepackage{comment}
\usepackage[margin=1in]{geometry}

\usepackage{amsmath, amssymb, amsthm, amsfonts}
\usepackage{color,cleveref}
\usepackage{graphicx,subcaption} 

\title{Viscous flow model of ion-induced pattern formation: consistency between theory and experiment}
\author[1]{Tyler P. Evans\footnote{Corresponding author: evans.tyler@utah.edu, https://orcid.org/0000-0001-7812-2479}}
\author[2]{Scott A. Norris}
\affil[1]{Department of Mathematics, University of Utah, Salt Lake City, UT 84112, United States of America}
\affil[2]{Department of Mathematics, Southern Methodist University, Dallas, TX 75275, United States of America}

\begin{document}
	
\maketitle

\begin{abstract}
    It is known that ion-irradiation can lead surfaces to spontaneously develop patterns with characteristic length scales on the order of only a few nanometers. Pattern formation typically occurs only for irradiation angles beyond a \textit{critical angle}, which varies with projectile, target, and irradiation energy. To date, there is no completely unifying physical theory. However, since predictions of the critical angle can be extracted from linear stability analysis of a given model, the ability to explain critical angle selection is a simple but important test of model validity.   
    In this paper, we survey all existing critical angle, wavelength, and in-plane stress data for noble gas broad beam ion-irradiation of silicon and germanium by argon, krypton and xenon at energies from 250eV to 2keV. While neglecting the effects of erosion and redistribution, which are widely regarded as key contributors to ion-induced pattern formation, we find that a viscous flow model is capable of explaining these three types of experimental data simultaneously using only parameters within a physically-reasonable range, and in a manner consistent with the defect kinetics of the amorphous layer. This bolsters the case for viscous flow and stress-driven instabilities as important components of an eventual, unifying model of nanoscale pattern formation under ion bombardment. 
\end{abstract}

\tableofcontents


\vspace{1cm}




\section{Introduction}


Observations of nanometer-scale self-organization of structures atop surfaces irradiated by broad ion beams have been documented since at least the 1960s \cite{navez-etal-1962}. These observations initiated a great deal of interest in the potential to exploit the phenomenon for commercial purposes, with possible applications in medical devices, semiconductor manufacturing, and other areas which would benefit from the ability to fine-tune the nanoscale self-organization process by toggling only coarse system parameters. If fully realized, this would enable the mass-production of useful nanostructures, such as nanowires, nanotubes, hexagonal arrays, and more, ``grown" via broad-beam irradiation, rather than costly engineering at lithographic and sub-lithographic scales. However, a unified theory capable of guiding engineering applications has proven remarkably elusive, and most applications at the time of writing rely on empirical, \textit{ad hoc} parameter tuning or machine learning methods \cite{chan-chason-JAP-2007,munoz-garcia-etal-MSER-2014,NorrisAziz_predictivemodel,cuerno-kim-JAP-2020-perspective}. Such methods sidestep the development of new physical models and suffer from difficulty in extrapolating to unexplored regions of parameter space.

Early theoretical work soon after the initial observations described above focused primarily on the sputter removal of atoms from the free interface \cite{sigmund-PR-1969,sigmund-JMS-1973}, which was shown to imply a morphological instability leading to pattern formation \cite{bradley-harper-JVST-1988,bradley-PRB-2011b}. However, this instability was predicted by theory to occur for all angles of incidence, in contrast to experimental observations of some ``critical angle" $\theta_c$ of incidence beyond which the instability occurs. This led to further work on modeling the redistribution of atoms \textit{not} sputtered away \cite{carter-vishnyakov-PRB-1996}, and, eventually, the Crater Function Framework (CFF) \cite{norris-etal-2009-JPCM}. Despite leading to improvements, the Crater Function Framework could not fully reconcile theory with experimental observations, and led to incorrect prediction of trends in $\theta_c$ \cite{norris-arXiv-2014-pycraters}. This has prompted a variety of \textit{ad hoc} modifications, some of which have been questioned \cite{hofsass-bobes-zhang-JAP-2016,NorrisAziz_predictivemodel}. 

Since around the late 1990s \cite{rudy-smirnov-NIMB-1999,umbach-etal-PRL-2001}, an alternative approach to understanding ion-induced self-organization has co-existed with the erosive-redistributive framework. Motivated by the extensive, radiation-induced damage of the crystalline structure, the top few nanometers of the irradiated target are treated as a highly viscous, amorphous thin film. Within the amorphous layer, stresses induced by ion implantation are relaxed by viscous flow, possibly leading to pattern formation \cite{cuerno-etal-NIMB-2011,castro-cuerno-ASS-2012,castro-etal-PRB-2012,norris-PRB-2012-linear-viscous,moreno-barrado-etal-PRB-2015,Swenson_2018,munoz-garcia-etal-PRB-2019,evans-norris-JPCM-2022,evans-norris-JPCM-2023} in a manner analogous to classical hydrodynamic instabilities \cite{drazin-reid-book,chandrasekhar-book-2013,cross-greenside-book}. Such models are sometimes referred to as \textit{hydrodynamic-type} models \cite{castro-cuerno-ASS-2012}.

Our group has previously developed a viscous flow model based on two sources of ion-induced stresses. First, we have considered Anisotropic Plastic Flow (APF) as a source of deviatoric stresses \cite{norris-PRB-2012-linear-viscous}, seeking analogy with the ion-hammering effect well-known at much higher energies \cite{trinkaus-ryazanov-PRL-1995-viscoelastic,trinkaus-NIMB-1998-viscoelastic,van-dillen-etal-PRB-2005-viscoelastic-model,wesch-wendler-book-2016}, despite different physics. Other work has also supported the use of such a model even at unexpectedly low energies \cite{van-dillen-etal-PRB-2005-viscoelastic-model,george-etal-JAP-2010}. APF, taken on a phenomenological basis, leads to an angle-dependent morphological instability \cite{norris-PRB-2012-linear-viscous} and compressive-to-tensile transition of in-plane stresses, agreeing with some experimental observations \cite{perkinsonthesis2017}. Second, we have considered a simple model of Ion-induced Isotropic Swelling (IIS), a source of isotropic stresses. IIS has been shown to stabilize the free surface of the film against perturbations, suppressing pattern formation \cite{Swenson_2018,evans-norris-JPCM-2022,evans-norris-JPCM-2023}, or, equivalently, increasing the value of $\theta_c$ \cite{evans-norris-JPCM-2023,evans-norris-JEM-2024}.


It is increasingly clear that both erosive-redistributive \textit{and} viscous flow mechanisms occur--- the questions are primarily of relative magnitude under different experimental conditions. Indeed, recent work suggests that the linear superposition of erosion-based and stress-based models offers the best explanation of experimental observations of 1keV Ar$^+$ on Si at low flux \cite{norris-etal-SREP-2017}. In the nonlinear regime of surface evolution, too, it has been suggested that viscous flow may play an important role alongside erosion \cite{myint-ludwig-etal-PRB-2021-Ar-bombardment,myint-ludwig-PRB-2021-Kr-bombardment}. Hence the development of a unifying theory has been stifled not only by the matter of cataloging all relevant mechanisms, but also of understanding their relative significance.

The large parameter space and mechanistic uncertainty suggest that theoretical advances could benefit from a data-driven approach. Accordingly, for the first time, we collect all known experimental data for critical angles $\theta_c$, the minimal incidence angle that leads to ripple formation of the irradiated surface. This data set comprises 28 experimental systems encompassing Ar$^+$, Kr$^+$, Xe$^+$ on Si and Kr$^+$, Xe$^+$ on Ge between 250eV and 2000eV. For the same ion-target-energy combinations, we also collect all known experimental data for wavelengths $\lambda(\theta,E,f)$ (48 data points), and in-plane stresses (20 data points).

In this paper, we demonstrate that a viscous flow model of ion-induced nanopatterning is capable of parsimoniously explaining all of the experimental data for $\theta_c$, $\lambda(\theta,E,f)$, and in-plane stresses, while respecting all observed experimental trends, using only physically-plausible parameter values, and in a manner consistent with recent work on the defect kinetics underlying the ion-enhanced fluidity \cite{ishii-thesis-2013,ishii-etal-JMR-2014}. Remarkably, we do so without invoking erosion and redistribution, the prevailing explanations for pattern formation. Hence our work represents a possible overturning of the \textit{status quo} theoretical description of low-energy pattern formation. Further implications of these findings are discussed in Section \ref{section:discussion}.


\section{Experimental data}


\label{section:data}
There has been a great deal of experimental work on the low-energy irradiation of Si and Ge with Ar$^+$, Kr$^+$, Xe$^+$ dating back as far as the 1960s \cite{navez-etal-1962}. Typical experiments consider the dependence of pattern formation on only one or two system parameters: most often ion or target species, energy, and angle of incidence, but sometimes ambient temperature, flux, fluence, and even mechanical vibration of the irradiated substrate. Commonly-recorded results are RMS roughness of the surface, angle-dependence wavelength $\lambda(\theta)$, and the critical angle $\theta_c$. Sometimes, in-plane stresses are also recorded. See Tables \ref{table:Si-data}-\ref{table:stresses-Ar-Si-offnormal} for a comprehensive summary of the experimental literature. Here, we will systematically make use of data on critical angles, wavelengths, and in-plane stresses.

In Tables \ref{table:Si-data} and \ref{table:Ge-data}, we summarize all existing data that we are aware of for $\theta_c$ from experimental systems involving Ar$^+$, Kr$^+$ and Xe$^+$ irradiation of Si or Ge targets at irradiation energies for which patterns form between 250eV and 2000eV for at least some angle of incidence. Ne$^+$ $\to$ Si is not considered, since \cite{frost-etal-APA-2008} reports that no patterns are formed in this energy range. Ne$^+$ and Ar$^+$ $\to$ Ge are also excluded due to lack of apparent pattern formation in this energy range \cite{frost-etal-APA-2008,Teichmann2013}. For these systems, it is unclear whether patterns fail to form due to mechanisms described within the viscous flow model described here (for example, IIS or BA are too strong, or APF is too weak), or for some other reason (for example, the viscous flow model is somehow inapplicable). A few aspects of the critical angle data are worth noting:
\begin{itemize}
    \item $\theta_c$ is generally higher for Ge targets than for Si targets. Per discussion in \cite{evans-norris-JPCM-2022,evans-norris-JPCM-2023}, this is consistent with the observation that Ge tends to produce more defects per ion, contributing to increased IIS.
    \item $\theta_c$ tends to be higher with increasing energy across all projectiles and targets.
    \item For each target, $\theta_c$ tends to be higher with heavier projectile species. This is exactly the opposite of what is expected if the instability-inducing mechanism were to scale simply with momentum, as is (approximately) the case in early versions of the Crater Function Framework \cite{norris-arXiv-2014-pycraters,NorrisAziz_predictivemodel}.
    \item As discussed in \cite{hofsass-bobes-zhang-JAP-2016}, the Ar$^+$ on Si data shows a sudden increase in $\theta_c$ near 1500eV. This has been plausibly explained in the context of a viscous flow model by \cite{evans-norris-JEM-2024}, where the effect of Boundary Amorphization is responsible.
    \item As discussed in \cite{Teichmann2013}, there is a significant increase in $\theta_c$ for Xe$^+$ on Ge between 250eV and 2000eV. Here, we speculate that this may represent the run-up to a sharp increase in $\theta_c$ for Xe$^+$ on Ge analogous to that of Ar$^+$ on Si near 1500eV. However, there is currently no experimental data available for Xe$^+$ on Ge between 2000eV and 5000eV.
\end{itemize}
It is also experimentally observed that $\lambda(\theta;E)$, for fixed $\theta$, projectile, and target, tends to scale with $\sqrt{E}$; see \cite{chini-etal-PRB-2003-TEM,ziberi-etal-JVSTA-2006,castro-etal-PRB-2012,hofsass-bobes-zhang-JAP-2016}. At low energies, normal incidence, and for fixed flux, the in-plane stress appears to scale like $E^{-7/6}$; see \cite{davis-TSF-1993-simple-compressive-stress,moreno-barrado-etal-PRB-2015,evans-norris-JEM-2024}.

\section{Continuum model}
\label{section:model}



\begin{figure}
    \centering
    \includegraphics[width=0.5\linewidth]{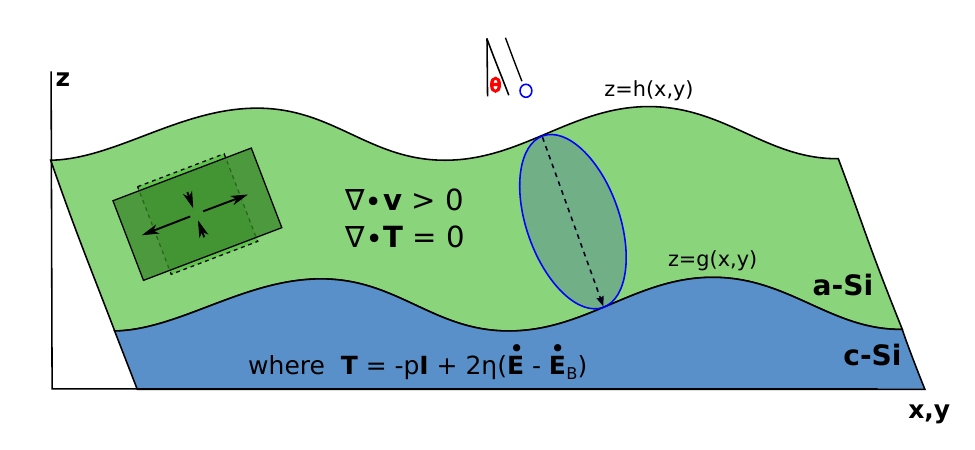}
    \caption{Schematic showing the convention for the irradiation angle $\theta$; a hypothetical collision cascade (blue ellipse); and the placement of upper (free) and lower (amorphous-crystalline) interfaces, $h(x,y)$ and $g(x,y)$, respectively. As an example, we consider a Si target.}
    \label{fig:schematic}
\end{figure}

As discussed in Section \ref{section:data}, the physics of irradiated surfaces is complex.  An incoming ion penetrates some characteristic distance into the target before initiating a  \textit{collision cascade} --- a sequence of successive recoils --- leading to numerous displacements of target atoms \cite{Kinchin-Pease-RoPP-1955} and the creation of defects \cite{chan-chason-JAP-2007}.  In semiconductors irradiated at room temperature, the accumulating defects cause a thin film of material at the surface to become \textit{amorphous}.  Meanwhile, some target atoms are sputtered away from the free surface, causing the gradual erosion of the target.  We here adopt a model that treats the irradiated amorphous film as a highly viscous fluid, with a fluidity $\eta^{-1}$ that is posited to be enhanced by the ion beam \cite{rudy-smirnov-NIMB-1999,umbach-etal-PRL-2001}. This hypothesis allows standard approaches from fluid dynamics to be employed \cite{drazin-reid-book,chandrasekhar-book-2013}. However, the complexity of the physics manifests in five unique deviations from the standard approach, which we summarize here (see \cite{evans-norris-JEM-2024} for more details) and depict schematically in Figure~\ref{fig:schematic}.

\subsection{Governing equations}

\paragraph{Bulk mass conservation with Ion-Induced Swelling.} 
Our model of the bulk amorphous layer begins with the standard equation of conservation of mass:
\begin{equation}
		\begin{gathered}
			\frac{\partial \rho}{\partial t} + \nabla \cdot (\rho \vec{v}) = 0, \label{consofmass}
		\end{gathered}
\end{equation}
where $\rho(x,z,t)$ is the density field, and $\vec{v}(x,z,t)$ is the velocity field.  Due to ion implantation and/or damage accumulation, the density of the film does not remain constant, but rather increases with exposure to the ion beam.  We model this behavior with a simple equation of state
 \begin{equation} \label{equationofstate}
		\rho = \frac{\rho_a}{1+\Delta}.
	\end{equation}
where $\rho_a$ is the baseline density of the amorphous target, and $\Delta (x,z,t)$ is a measure of the accumulated swelling of a material parcel, itself governed by an advection equation with constant forcing:
	\begin{equation}
		\frac{\partial \Delta}{\partial t} + \vec{v} \cdot \nabla \Delta = f A_I, \label{advectioneqn}
	\end{equation}
Together, Equations (\ref{equationofstate}) and (\ref{advectioneqn}) form a simple model of Ion-Induced Swelling (IIS).  They simply suppose that, to leading order, changes in density due to irradiation are dominated by accumulated damage to the crystalline target, with negligible dependence on the pressure.  This choice is made, in part, in order to avoid turning the present work into a referendum on equations of state for amorphous matter, an ongoing area of study \cite{guo-li-JAP-2010}. Despite its simplicity, the above form has been successfully used in modeling of experimental results elsewhere \cite{li-etal-vacuum-2024}.  Finally, we note that Equations (\ref{consofmass}) - (\ref{advectioneqn}) could be rewritten in terms of the divergence of the velocity field,
    \begin{equation}
        \nabla\cdot \vec{v} = \frac{\rho}{\rho_a}fA_I; \label{eqncombined}
    \end{equation}
this presentation emphasizes that the chief effect of IIS is to reduce density while conserving mass, leading to ``volumization".

\paragraph{Bulk momentum conservation with Anisotropic Plastic Flow (APF).}
We next introduce the conservation of linear momentum, which, in the absence of body forces, is
\begin{equation}
		\rho \left( \frac{\partial \vec{v}}{\partial t} + \vec{v} \cdot \nabla \vec{v} \right) = \nabla \cdot \textbf{T},
\end{equation}
where $\textbf{T}$ is the Cauchy stress tensor. In the limit of small Reynolds number \cite{drazin-reid-book}, valid here due to small film thickness and extremely high viscosity, the above is simplified to
\begin{equation}
    \nabla \cdot \textbf{T} = 0.
\end{equation}
Following discussion in \cite{norris-PRB-2012-linear-viscous}, we adopt the Cauchy stress tensor
 \begin{equation}
 			\textbf{T} = - p\textbf{I} + 2\eta(\dot{\textbf{E}} - \dot{\textbf{E}}_b),
        \end{equation}
with $\dot{\textbf{E}} = \frac{1}{2}\left( \nabla \vec{v} + \nabla \vec{v}^T \right)$ the linear strain rate tensor. In the above, $p$ is pressure, $\eta$ is the viscosity, and $\textbf{I}$ is the identity tensor.  In addition, the effect of the ion-beam is incorporated through the extra term
        \begin{equation} \label{tensordef}
			\dot{\textbf{E}}_b = f A_D \mathbf{D} (\theta) 
                \equiv f A_D
			\begin{bmatrix} 
				\frac{3}{2}\cos(2\theta) - \frac{1}{2} & 0 & \frac{3}{2}\sin(2\theta) \\
				0 & 1 & 0 \\
				\frac{3}{2}\sin(2\theta) & 0 & -\frac{3}{2}\cos(2\theta) - \frac{1}{2} \\
			\end{bmatrix} 
	\end{equation}
 which represents \emph{Anisotropic Plastic Flow} (APF); see \cite{van-dillen-etal-PRB-2005-viscoelastic-model,otani-etal-JAP-2006}. This term is taken on a phenomenological basis, since the irradiation energies of interest in the present work are too small for the mechanisms discussed in \cite{trinkaus-NIMB-1998-viscoelastic,van-dillen-etal-APL-2001-colloidal-ellipsoids,van-dillen-etal-APL-2003-colloidal-ellipsoids,van-dillen-etal-PRB-2005-viscoelastic-model,vandillen-etal-prb-2006,wesch-wendler-book-2016} to be active. However, the mathematical form of this stress tensor has still been successfully used in modeling even at low energies \cite{van-dillen-etal-PRB-2005-viscoelastic-model,george-etal-JAP-2010,norris-PRB-2012-linear-viscous}. Here, $A_D$ is the deformation rate associated with APF, and $f$ is the nominal flux. It is shown elsewhere \cite{norris-PRB-2012-linear-viscous} that this effect leads to an angle-dependent morphological instability.

\paragraph{Top boundary conditions, including sputtering.} 
At the free upper interface, $z=h(x,t)$, we have kinematic and stress-balance conditions of the form
	\begin{equation}
		\begin{gathered}
			v_{I} = \vec{v}\cdot \hat{\mathbf{n}} - V (\theta) \frac{\rho_c}{\rho} \\
			\textbf{[[T]]} \cdot \hat{\mathbf{n}} = -\gamma \kappa \hat{\mathbf{n}},
		\end{gathered}
	\end{equation}
In the first of these equations, $v_I$ is the (normal) interfacial velocity, $\vec{v}\cdot \hat{\mathbf{n}}$ is the normal component of the fluid velocity, and the last term is a correction containing a downward velocity $V(\theta)$ that accounts for sputter erosion \cite{Swenson_2018}.  In the second of these equations, $\textbf{[[T]]} \cdot \hat{\mathbf{n}}$ is the jump in stress across the amorphous-vacuum boundary, $\gamma$ is surface energy and $\kappa$ is the mean curvature.

\paragraph{Bottom boundary conditions, including amorphization.}
At the lower, amorphous-crystalline interface, $z=g(x,t)$,
     \begin{equation} 
        [[\rho \vec{v}]]\cdot \hat{\mathbf{n}} = [[\rho]] v_{I} \label{eq: amorphization},
    \end{equation}
which is a statement of mass conservation across a possibly non-planar boundary \cite{davis-solidification-book-2001,evans-norris-JEM-2024}. If the densities on each side of the amorphous-crystalline boundary were the same, this would simplify to the usual no-penetration boundary condition. Since $\rho_a \neq \rho_c$ in general for irradiated semiconductors, and sputtering induces downward translation of the amorphous layer into the crystalline substrate, Equation (\ref{eq: amorphization}) introduces an important effect: that of phase-change at the moving interface. Throughout this work, we refer to this effect as Boundary Amorphization (BA) for convenience. Elsewhere, BA has been suggested as an explanation for the sudden increase of $\theta_c$ for Ar$^+ \to$ Si around 1.5keV irradiation \cite{evans-norris-JEM-2024}. In other contexts, phase change at moving boundaries may also significantly alter linear stability \cite{davis-solidification-book-2001}.

Next, the no-slip condition,
    \begin{equation} 
        \vec{v}\cdot\hat{\mathbf{t}} = 0, \label{no-slip}
    \end{equation}
is assumed. Here, $\hat{\mathbf{t}}$ is the unit tangent vector from the amorphous-crystalline interface. Although slip flow could possibly affect $\theta_c$, it would take unrealistically large slip lengths in order to achieve a difference of more than one or two degrees in $\theta_c$ \cite{evans-thesis}; it is therefore neglected in this work.

\paragraph{Relationship between boundaries.}
A final notable feature of ion-irradiated films is that the upper and lower interfaces are not independent.  Because the amorphous film is created by ions penetrating some distance into the film, and as the top interface recedes due to sputtering, the ions will reach further into the film causing it to recede as well.  Although the exact relationship between the two surfaces has been modeled in various ways (flat \cite{umbach-etal-PRL-2001,castro-cuerno-ASS-2012,munoz-garcia-etal-PRB-2019}, vertically-displaced \cite{norris-PRB-2012-viscoelastic-normal,norris-PRB-2012-linear-viscous}, or diagonally-displaced \cite{moreno-barrado-etal-PRB-2015,Swenson_2018}), our group has recently demonstrated the necessity of connecting the lower-interface geometry directly to collision cascade parameters \cite{evans-norris-JEM-2024}. This is achieved by defining the lower interface $g(x,t)$ in terms of the upper interface $h(x,t)$ as
\begin{equation}
    g(x,t) = h(x-x_0(\theta),t) - h_0(\theta)
\end{equation}
where
\begin{equation}
    \begin{gathered}
        x_0(\theta) = a\sin(\theta) + 2\left(\frac{(\alpha^2-\beta^2)\sin(\theta)\cos(\theta)}{\sqrt{\alpha^2\cos^2(\theta) + \beta^2\sin^2(\theta)}}\right), \\
        h_0(\theta) = a\cos(\theta) + 2\left(\sqrt{\alpha^2\cos^2(\theta) + \beta^2\sin^2(\theta)}\right).
    \end{gathered} \label{CCreln}
\end{equation}
Here, $a$ denotes the mean depth of ion implantation, $\alpha$ the downbeam standard deviation, and $\beta$ the crossbeam standard deviation. These quantities assume that the distribution of implanted ions follows a bivariate Gaussian ellipsoid \cite{sigmund-PR-1969,sigmund-JMS-1973} aligned with the ion beam, which is ``dragged" along the top interface to trace out the bottom one.  Notably, these quantities are readily obtained from simulations using either molecular dynamics, or the much faster Binary Collision Approximation (BCA), using software such as SRIM or one of its many descendants  \cite{ziegler-biersack-littmark-1985-SRIM,srim-2000.40}. 

\subsection{Linear stability results}



Previous analysis \cite{evans-norris-JPCM-2023,evans-norris-JEM-2024} of Equations (\ref{consofmass})-(\ref{no-slip}) reveals a steady shear flow with velocity profile
\begin{equation}
   \vec{v}_0 (z) = <3fA_D\sin(2\theta)z, \hspace{.25cm} V\bigg(\frac{\rho_c}{\rho_a}\bigg)\sqrt{1 + 2fA_I\frac{\rho_a}{\rho_c}\frac{z}{V}}>
\end{equation}
After linearizing about this steady state, the resulting equations are too complex to solve in closed form, so we employ several simplifying limits:  
\begin{itemize}
    \item the \textit{small perturbation limit}, which is typical of linear stability analysis, wherein we suppose that the free interface is perturbed weakly: $h(x,t) = h_0(\theta) + \epsilon h_1(x,t)$, where $0 < \epsilon \ll 1$;
    \item the \textit{long-wave limit}, wherein the perturbative wavenumber $k$ is vanishingly small compared to the film thickness, $h_0(\theta)$--- or, equivalently, $(kh_0(\theta)) \ll 1$. Indeed, pattern formation in noble gas ion-irradiated semiconductors appears to proceed by a long-wave (or ``Type II" \cite{cross-hohenberg-RoMP-1993,cross-greenside-book}) instability \cite{NorrisAziz_predictivemodel}; and
    \item the \textit{small-swelling} limit, $\frac{fA_Ih_0(\theta)}{V(\theta)} \ll 1$, which says that density changes in the bulk are small by the time an affected parcel of matter is sputtered away from the free interface \cite{Swenson_2018,evans-norris-JPCM-2022}.
\end{itemize}
To leading order in these limits, the governing equations can be solved \cite{evans-norris-JPCM-2023,evans-norris-JEM-2024}, yielding a dispersion relation $\sigma$ with real part

\begin{equation}
	\begin{aligned}
		\text{Re}\left( \sigma \right) = & - f \Bigg[ 
                   3 A_D\left(\frac{\rho_a}{\rho_c}\right) \frac{\cos\left( 2\theta + \Psi(\theta)\right)}{\cos\left( \Psi(\theta)\right)} \\
                   &+ \frac{A_I}{2}\left(\frac{\rho_a}{\rho_c}\right)^2 
                   + \frac{\Omega Y(\theta)}{h_0(\theta)}\left(1 - \frac{\rho_a}{\rho_c}\right)			          
               \Bigg] \big( kh_0(\theta)\big)^2 \\
               &- \frac{\gamma}{3\eta(\theta) h_0(\theta)} \left( k h_0(\theta)\right)^4 + \dots 
    \end{aligned} \label{dispnrelnwBA}
\end{equation}
where 
 \begin{equation}
    \Psi(\theta) = \arctan \left( \frac{x_0 (\theta)}{h_0 (\theta) } \right).  \label{eq: effective-displacement-angle}
\end{equation}

\subsection{Predictions on patterns, wavelengths, and stresses}



According to pattern-formation theory \cite{cross-hohenberg-RoMP-1993,cross-greenside-book}, we can expect to observe structures if $\text{Re} \left[ \sigma \left( k \right) \right] > 0$ for any wavenumber $k$.  In Equation (\ref{dispnrelnwBA}), this will occur for $k$ near zero if the coefficient of $k^2$ is positive, a condition which can be written
\begin{equation}
    - 3 A_D \frac{\cos\left( 2\theta + \Psi(\theta)\right)}{\cos\left( \Psi(\theta)\right)} >   \frac{A_I}{2}\left(\frac{\rho_a}{\rho_c}\right)
                   + \frac{\Omega Y(\theta)}{h_0(\theta)}
                   \left( \frac{\rho_c}{\rho_a} - 1 \right) 	;
\end{equation}
This equation illustrates that patterns occur when the influence of Anisotropic Plastic Flow both (a) becomes destabilizing above a critical angle $\theta_C$, and (b) exceeds the combined stabilizing influences of Irradiation Induced Swelling and Boundary Amorphization.  

If patterns form, then at early times (i.e. within the linear regime where this analysis is valid) we expect their wavelength to correspond to the wavenumber $k^*$ that maximizes $\text{Re} \left[ \Sigma \left( k \right) \right]$, which can also be computed from Equation (\ref{dispnrelnwBA}):
\begin{equation}
    \lambda^*(\theta) = \sqrt{\frac{2}{3}}\pi h_0(\theta) 
    \sqrt{ 
        \frac{
            \gamma \eta^{-1}(\theta)
        }{
            - 3 f A_D h_0(\theta)\frac{\rho_a}{\rho_c}\left( \frac{\cos\left( 2\theta + \Psi(\theta)\right)}{\cos\left( \Psi(\theta)\right)}\right) 
            - \frac{1}{2}f A_I h_0(\theta)\left( \frac{\rho_a}{\rho_c} \right)^2
            - f \Omega Y(\theta) \left( 1 - \frac{\rho_a}{\rho_c} \right) 
        }
    }, \label{wavelength-prediction}
\end{equation}
where $E,f,\Omega,\gamma, \rho_a$ and $\rho_c$ are fixed. Finally, the steady in-plane stress is \cite{evans-norris-JPCM-2023}
\begin{equation}
    \textbf{T}_{0xx} = 6fA_D\eta(\theta)\cos(2\theta) + 2fA_I\eta(\theta). \label{steady-stress}
\end{equation}

\subsection{Important notes on fluidity, $\eta^{-1}(\theta)$}
\label{subsection:fluidity}


As can be seen in Equations (\ref{wavelength-prediction}) and (\ref{steady-stress}), theoretical predictions on the steady stress $|\textbf{T}_{xx}|$ and ripple wavelength $\lambda(\theta)$ are possible only with knowledge of the ion-enhanced fluidity $\eta^{-1}(\theta)$. Unfortunately, this parameter is essentially impossible to measure directly, and also very difficult to estimate. It is also evidently flux-dependent \cite{NorrisAziz_predictivemodel,vauth-mayr-PRB-2007,vauth-mayr-PRB-2008b,madi-etal-PRL-2011,norris-etal-NCOMM-2011}. 
Given the crucial role that $\eta^{-1}(\theta)$ plays in understanding these experimental systems, some attempts have been made to develop reasonable approximations and simple models. In \cite{norris-etal-SREP-2017}, it has been suggested that
\begin{equation}
    \eta^{-1}(\theta,E,f) \propto f\cos(\theta)\frac{E(\theta)}{h_0(\theta)}, \label{fluidity-VM}
\end{equation}
which implies that a scaling is possible if a $\eta^{-1}(\theta,E,f)$ is known for a fixed angle, energy, and flux:
\begin{equation}
    \frac{\eta^{-1}(\theta,E,f)}{\eta^{-1}(0,1000 \text{ eV},10 \frac{\text{ions}}{\text{nm}^2\cdot\text{s}} )} = \frac{f}{ 10\frac{\text{ions}}{\text{nm}^2\cdot\text{s}}}\cdot \frac{\cos(\theta) E(\theta)}{(1000 \text{eV})}\cdot \frac{h_0(0, 1000 \text{eV} )}{h_0(\theta, E)}.\label{scaling-VM}
\end{equation}
However, the prevailing arguments \cite{norris-etal-NCOMM-2011,madi-etal-2008-PRL,hofsass-bobes-zhang-JAP-2016} in favor of the form (\ref{fluidity-VM}) rely on a single set of molecular dynamics (MD) experiments performed at approximately room temperature \cite{vauth-mayr-PRB-2007,vauth-mayr-PRB-2008b}. It has been suggested that these results may break down with increased energy; see discussion in the Supplemental Material of \cite{norris-etal-NCOMM-2011}. Throughout this work, we will refer to this model as the \textit{Vauth-Mayr model} of fluidity, or \textit{VM} for conciseness, while acknowledging that \cite{vauth-mayr-PRB-2007,vauth-mayr-PRB-2008b} did not claim the validity of (\ref{scaling-VM}); rather, other groups took inspiration from their work.

Separately, a model based on the ion-induced creation and recombination of defects \cite{ishii-etal-JMR-2014} has proven necessary to explain the apparent ``freezing-in" of stress when irradiation is ceased, and the same model appears to predict a different scaling of $\eta^{-1}(\theta)$ with flux $f$ and nominal energy $E$. In \cite{ishii-etal-JMR-2014}, it is argued that
\begin{equation}
    \eta^{-1}(\theta,E,f) = \alpha \bigg[\frac{f\cos(\theta)E(\theta)}{ D_2 h_0(\theta,E)}\bigg]^{1/2}, \label{fluidity-ishii}
\end{equation}
where we have brought their notation into alignment with that of the present work and made use of the fact that defects are proportional to absorbed energy\footnote{This implies that the bimolecular annihilation rate, $D_2$, and the fluidity-per-defect $\alpha$ take on a slightly different meaning here than in \cite{ishii-thesis-2013,ishii-etal-JMR-2014}.} \cite{Kinchin-Pease-RoPP-1955}. If we assume, as \cite{ishii-etal-JMR-2014} did, that $D_2$ and $\alpha$ are constants with respect to angle, flux, and energy, then this implies an alternative scaling,
\begin{equation}
    \frac{\eta^{-1}(\theta,E,f)}{\eta^{-1}(0,1000 \text{ eV}, 10 \frac{\text{ions}}{\text{nm}^2 \text{s}})} = \sqrt{\frac{f\cos(\theta)}{10 \frac{\text{ions}}{ \text{nm}^2\text{s}}}}\sqrt{\frac{E(\theta)}{1000 \text{eV}} \frac{h_0(0, 1000 \text{eV})}{h_0(\theta, E)}}. \label{scaling-Ishii}
\end{equation}
We will refer to this as the \textit{Ishii model} of fluidity, or \textit{Ishii} for conciseness, while acknowledging that the ion-induced stress, rather than the fluidity \textit{per se}, was of primary interest in \cite{ishii-thesis-2013,ishii-etal-JMR-2014}.
In what follows, we take $E(\theta) = E(0)$
--- that is, the explicit angular dependence of absorbed energy is negligible --- for simplicity, and in order to avoid needing to explicitly calculate the absorbed energy as a function of angle from BCA software. We anticipate that this simplification is permissible, as the difference between $E(\theta)$ and $E(0)$ is typically quite small (on the order of 10\%) far from grazing incidence. The majority of our data for $\lambda(\theta,E,f)$ and $T_{0xx}$ is also far from grazing incidence, which makes this simplification broadly appropriate; see Tables \ref{table:wavelengths-Si-Ar}-\ref{table:stresses-Ar-Si-offnormal}.


\section{Comparing theory and experiment}
\label{section:results}

As noted above, a primary barrier to the development of a predictive model for ion-induced pattern formation is the large number of parameters.  In Equation~\eqref{dispnrelnwBA}, these include
\begin{itemize}
\item  \textbf{experimental} parameters:  $\{ Z_T, Z_I, E, \theta, f \}$
\item  \textbf{atomistic} parameters:  $\{ a, \alpha, \beta, Y, h_0, x_0 \}$
\item  \textbf{continuum} parameters:  $\{A_D, A_I, \left( 1 - \tfrac{\rho_a}{\rho_c} \right), \eta^{-1} \}$
\end{itemize}
where, in general, any of the atomistic or continuum parameters can depend on any or all of the experimental parameters.  Now, for any given data point, the experimental parameters are known, and the atomistic parameters can be estimated using simulation software such as SRIM \cite{ziegler-biersack-littmark-1985-SRIM,srim-2000.40,evans-norris-JEM-2024}.  However, each of the main \emph{mechanistic coefficients} -- (a) the coefficient $A_D$ of the (deviatoric) Anisotropic Plastic Flow, (b) the coefficient $A_I$ of the (isotropic) Irradiation-Induced Swelling, (c) the relative density change $\left( 1 - \tfrac{\rho_a}{\rho_c} \right)$, and (d) the ion-induced fluidity $\eta^{-1}$ -- is unknown, and difficult to measure directly.  In general, therefore, fitting must be performed.

For any given combination of experimental parameters, there are at most two measured values: the characteristic pattern wavelength (if present), and the steady stress.  Moreover, only rarely are both of these values available for the same system.  Thus, if all continuum coefficients are free to fit, then the system is undetermined, in particular because the critical angle and wavelength are invariant to the scalar multiplication of all four parameters by the same constant. Therefore, we take three measures to manage the number of degrees of freedom in the system.

\paragraph*{Non-dimensionalization.}
First, we non-dimensionalize our dispersion relation to obtain at least one term without a coefficient.  Because we are studying the formation of patterns, which requires $A_D > 0$, we use as a scaling factor the quantity $fA_D\frac{\rho_a}{\rho_c}$.  Dividing by this quantity, we obtain
\begin{equation}
			\text{Re}\left( \Sigma \right) = \Bigg[ - 3 \frac{\cos\left( 2\theta + \Psi(\theta)\right)}{\cos\left( \Psi(\theta)\right)} - \frac{1}{2} P_1 - P_2 \Bigg] \left( kh_0\right)^2 - P_3 (kh_0)^4,
        \label{simple-dispnrelnwBA}
\end{equation} 
where
\begin{equation}
     \Sigma = \frac{\sigma}{fA_D} \frac{\rho_c}{\rho_a}, \hspace{.25cm}
        P_1 = \frac{A_I}{A_D} \frac{\rho_a}{\rho_c}, \hspace{.25cm}
        P_2 = \frac{\Omega Y}{A_Dh_0} \left( \frac{\rho_c}{\rho_a} - 1 \right), \hspace{.25cm}
        P_3 = \frac{\gamma \eta^{-1}}{3fA_D h_0}\frac{\rho_c}{\rho_a}
        \hspace{.25cm}
     \label{eq: nondim-definitions-2}
\end{equation}

\paragraph*{Functional forms.}
Second, we will make the following assumptions on the parametric forms of the continuum coefficients.  For each ion/target combination, we will assume that
\begin{equation}
     \begin{aligned}
        \frac{ \rho_a (Z_T, Z_I, E, \theta, f) }{ \rho_a (1000, 0, 10) } &= 1 \\
        \frac{ A_I (Z_T, Z_I, E, \theta, f) }{ A_I (Z_T, Z_I, 1000, 0, 10) } &= \left( \frac{E}{1000} \right)^{c_1} \\
        \frac{ A_D (Z_T, Z_I, E, \theta, f) }{ A_D (Z_T, Z_I, 1000, 0, 10) } &= \left( \frac{E}{1000} \right)^{c_2} \\
         \frac{ \eta^{-1} (Z_T, Z_I, E, \theta, f) }{ \eta^{-1} (Z_T, Z_I, 1000, 0, 10) } &= \left( \frac{E}{1000} \frac{f\cos(\theta)}{10} \frac{h_0(0,1000)}{h_0(\theta,E)}  \right)^{c_3}.
     \end{aligned}
     \label{eq: parameter-scaling-assumptions}
\end{equation}
That is, (a) the amorphous density -- and hence the relative density change -- is constant across energies, angles, and fluxes; (b) the single-impact distortion coefficients scale with energy, but not with flux; and (c) the fluidity scales with both energy and flux, according to discussion in Section \ref{subsection:fluidity}. Assumption (a) is equivalent to $\rho_a$ being a material-dependent constant. Assumption (b) is supported by observations due to \cite{van-dillen-etal-APL-2001-colloidal-ellipsoids,van-dillen-etal-APL-2003-colloidal-ellipsoids,van-dillen-etal-PRB-2005-viscoelastic-model} and others. Assumption (c) is generally agreed-upon \cite{chan-chason-JAP-2007,hofsass-bobes-zhang-JAP-2016,hofsass-bobes-APR-2019,NorrisAziz_predictivemodel}, and well-supported by existing experimental-theoretical study \cite{ishii-thesis-2013,ishii-etal-JMR-2014,norris-etal-SREP-2017}. Combining equations \eqref{eq: nondim-definitions-2}-\eqref{eq: parameter-scaling-assumptions} allows us to re-write our dimensionless parameters in the form
\begin{equation}
    \begin{aligned}
        P_1  &= P_{10} \left( Z_T, Z_I \right) \left( \frac{E}{1000} \right)^{c_1-c_2} \\
        P_2  &= P_{20} \left( Z_T, Z_I \right) \left( \frac{E}{1000} \right)^{-c_2}  \left[ \frac{ Y(E, \theta)}{Y (1000, 0)} \frac{ h_0(1000, 0)}{h_0 (E, \theta)} \right] \\
        P_3  &= P_{30} \left( Z_T, Z_I \right) \left( \frac{E}{1000} \right)^{c_3-c_2}\left( \frac{f\cos(\theta)}{10} \right)^{c_3} \left( \frac{f}{10} \right)^{-1} \left[ \frac{ h_0(1000, 0)}{h_0 (E, \theta)} \right]^{c_3+1} \label{scalings}
    \end{aligned}
\end{equation}
With these steps, we have reduced the number of parameters to three \emph{per ion/target combination}, plus an extra three that are expected to hold across all such combinations.  Thus, over-determination can be achieved if we have at least $3N+3$ data points across $N$ ion-target combinations, a condition which is definitely present in the composite data set described in Section~\ref{section:data}.


\paragraph*{Sequential fitting.}
We will compare our model to the experimental data \emph{sequentially}:
\begin{itemize}
    \item First, we seek agreement between experimental $\theta_c$ for Ar$^+$, Kr$^+$ and Xe$^+$ irradiated Si and Ge. This requires only the quadratic coefficient of Equation (\ref{simple-dispnrelnwBA}) and is the point of comparison between theory and experiment requiring the fewest free parameters.
    \item Second, having obtained good agreement between our model and $\theta_c$, we append the quartic term and seek agreement between theoretical wavelength predictions from (\ref{wavelength-prediction}) and the experimental data $\lambda(\theta)$. This now requires knowledge of $\eta^{-1}(\theta)$, one extra fit parameter, and affords us the opportunity to compare the predictions made by the two existing approaches to modeling $\eta^{-1}(\theta)$.
    \item Third, having obtained good agreement with $\theta_c$ and $\lambda(\theta)$ for a variety of systems, we return to dimensional quantities. Using Equation (\ref{steady-stress}), we fit the existing data for the in-plane stresses of Ar$^+$-irradiated Si using $\frac{\rho_a}{\rho_c}$ as the only remaining free parameter. This acts as a final validation step for the model: it is, in principle, possible to obtain good fits in the first two stages using dimensionless parameter combinations while relying on unreasonable \textit{physical} parameter values. Because $\frac{\rho_a}{\rho_c}$ has been estimated by other means and must lie in the range $(0,1]$, obtaining a nonphysical fitted value leads immediately implies the nonphysicality of our model. On the other hand, \textit{successful} fitting here produces a fully fitted model capable of simultaneously explaining $\theta_c$, $\lambda(\theta)$ and $\textbf{T}_{0xx}$. We again use the stress data to compare the two existing hypotheses on $\eta^{-1}(\theta)$, where the favorability of the form (\ref{fluidity-ishii}) is even more stark.
\end{itemize}

\subsection{Comparison to critical angles}

In previous work \cite{castro-cuerno-ASS-2012,castro-etal-PRB-2012,moreno-barrado-etal-PRB-2015,evans-norris-JEM-2024}, subsets of data for $\theta_c$ and $\lambda(\theta)$ are compared with theoretical models. In some cases, reasonable agreement with $\theta_c$ or $\lambda(\theta)$ can be obtained using a given model and a small number of fit parameters. Here we have collected all known data for $\theta_c, \lambda(\theta)$ and in-plane stresses for Ar$^+$, Kr$^+$, and Xe$^+$ on Si and Ge targets at energies between 250eV and 2000eV. Using the model described in Section \ref{section:model}, we will attempt to fit this much larger data set while maintaining a minimal number of free parameters. To this end, we compare the roots, $\theta_c$, of 
\begin{equation}
- 3 \frac{\cos\left( 2\theta + \Psi(\theta)\right)}{\cos\left( \Psi(\theta)\right)} - \frac{1}{2} P_1 - P_2
\end{equation} 
with the experimental data for $\theta_c$ outlined in Tables \ref{table:Si-data} and \ref{table:Ge-data}. The quantities $P_1$ and $P_2$ are as described in Equations (\ref{scalings}). $P_{10}$ and $P_{20}$ are specific to a given ion-target combination, while $c_1$ and $c_2$ are global parameters.


\paragraph{Fit 1: APF only.} As a na\"{i}ve baseline attempt, we first try to fit to a model containing only APF. As noted in \cite{norris-PRB-2012-linear-viscous,evans-norris-JPCM-2023,evans-norris-JEM-2024}, APF in the absence of IIS and BA selects $\theta_c$ without any dependence on its coefficient in the linear dispersion relation. The fits are very poor, as such a model cannot possibly account for the non-universality of $\theta_c$. For all systems, $\theta_c \approx 36^{\circ}$ is predicted.

\paragraph{Fit 2: APF + IIS.} We consider a model containing APF and IIS. We obtain reasonably good fits for Ge targets only because $\theta_c$ for Kr$^+$ and Xe$^+$ irradiation is approximately linear in $E$. However, this model is incapable of explaining the sudden increase in $\theta_c$ for Ar$^+$ $\to$ Si near 1500eV. 

\paragraph{Fit 3: APF + BA.} We consider a model containing BA as the sole angle-independent stabilizing mechanism and exclude IIS entirely. The fitted model does not respect the trends within the $\theta_c$ vs $E$ data, and the fits are generally of poor quality. We conclude that IIS should be present in the model.

\paragraph{Fit 4: APF + IIS + BA (I).} A model containing all of the mechanisms present in \cite{evans-norris-JEM-2024} is now used. The fits are acceptable, and the model is capable of fitting both the near-linear energy dependence of $\theta_c$ for Kr$^+$, Xe$^+$ on Si and Kr$^+$, Xe$^+$ on Ge, as well as the sudden increase in $\theta_c$ around 1500eV. From the fitted values of $c_1$ and $c_2$, we find that APF and IIS have indistinguishable energy scalings. This suggests that APF and IIS share a common cause, as might be expected, and prompts the elimination of one free variable.

\begin{figure}[htpb]
	\centering	
 \includegraphics[totalheight=7cm]{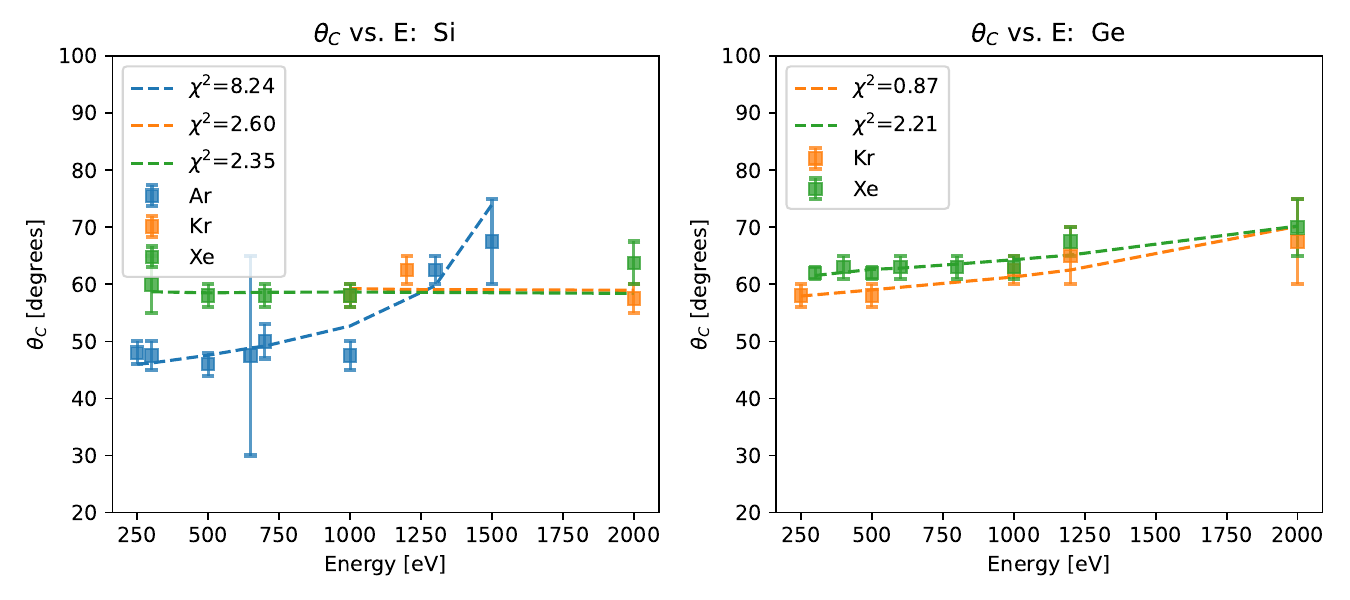}
 \caption{The results of fitting with APF, IIS and BA. One energy scale is shared between both $A_D$ and $A_I$: $c_1=c_2$.} \label{fig:all-onescale}
\end{figure}

\paragraph{Fit 5: APF + IIS + BA (II).} A second model containing all three mechanisms is considered, and the results are shown in Figure \ref{fig:all-onescale}. Based on the results of Fit 4, showing that $c_1$ and $c_2$ are indistinguishable, we here impose $c_1=c_2$, eliminating one fit parameter. This results is excellent fits and forms the basis for further comparison of our model with $\lambda(\theta)$ in Section \ref{section:discussion}.

\paragraph{Summary of results.} The results of the hierarchical fitting are summarized in Tables \ref{table:results-a} and \ref{table:results-b}. Of the models considered, the best is the one that includes all APF, IIS and BA, with APF and IIS having the same scaling with energy. 

\begin{table}[htpb]
\centering
\begin{tabular}{||c c c||} 
 \hline
Model & $\tilde{c}$ & $\chi_{\text{red}}^2$ \\ [0.5ex] 
 \hline\hline
 APF only & --- & 147.412 \\ 
 APF + IIS & 0.07 & 1.5834 \\
 APF + BA & -0.03 & 68.167 \\
 \hline
\end{tabular}
\caption{Summary of fitting results across models and for global fit parameter $\tilde{c}$, which represents the single free parameter in the model given.}
\label{table:results-a}
\end{table}

\begin{table}[htpb]
\centering
\begin{tabular}{||c c c c||} 
 \hline
Model & $c_1$ & $c_2$ & $\chi_{\text{red}}^2$ \\ [0.5ex] 
 \hline\hline
 APF + IIS + BA (I) & 0.03 & 1.49 & 0.611 \\
 APF + IIS + BA (II) & -\textbf{1.32} & -1.32 & 0.593 \\ [1ex] 
 \hline
\end{tabular}
\caption{Summary of fitting results across models and for global fit parameters $c_1$ and $c_2$. In the last row, $c_1$ is set to $c_2$ and is not an independently-fittable parameter. }
\label{table:results-b}
\end{table}

\begin{table}[htpb]
\centering
\begin{tabular}{||c c c c||} 
 \hline
Ion & Target & $P_{10}$ & $P_{20}$ \\ [0.5ex] 
 \hline\hline
 Ar & Si & 3.123 & 0.291 \\ 
 Kr & Si & 7.058 & 0 \\ 
 Xe & Si & 7.129 & 0 \\ 
 Kr & Ge & 5.922 & 0.095 \\ 
 Xe & Ge & 6.995 & 0.018 \\ 
 [1ex] 
 \hline
\end{tabular}
\caption{Summary of ion-target specific fit parameters, $P_{10}$ and $P_{20}$. Model ``APF + IIS + BA (II)" is assumed.}
\label{table:results-thetac}
\end{table}


\subsection{Comparison to wavelengths}
\label{subsection:wavelengths}

According to the discussion above, the fastest-growing wavelength $\lambda(\theta)$ is
\begin{equation}
    \lambda(\theta) = 2\pi\sqrt{\frac{-P_3h_0^2(\theta)}{3 \frac{\cos\left( 2\theta + \Psi(\theta)\right)}{\cos\left( \Psi(\theta)\right)} + \frac{1}{2} P_1 + P_2}}, \label{wavelengths}
\end{equation}
which permits comparison with experimental wavelength data. Since $P_1$ are $P_2$ are fully determined from fitting to $\theta_c$ (see Table \ref{table:results-thetac}), $P_3$ is the only component of the model available to fit the model to $\lambda(\theta)$. We adopt the form for $P_3$ described in (\ref{scalings}). There, the form (\ref{fluidity-VM}) is implemented by setting $c_3=1$, and (\ref{fluidity-ishii}) by setting $c_3=.5$. Then $P_{30}$ is the only fit parameter allowed for each ion-target combination. 


For some of these systems, $\lambda(\theta)$ is available only for a few $\theta$, or wavelengths are highly uncertain. Sometimes, the flux reported in the source literature is given as a range; in these cases, we use the midpoint of the given range. See Tables \ref{table:wavelengths-Si-Ar}, \ref{table:wavelengths-Si-Kr}, \ref{table:wavelengths-Si-Xe}, \ref{table:wavelengths-Ge-Kr}, and \ref{table:wavelengths-Ge-Xe}. We also note that (\ref{wavelengths}) is a long-wave approximation. As discussed elsewhere \cite{norris-PRB-2012-linear-viscous,NorrisAziz_predictivemodel}, this approximation leads to somewhat different theoretical predictions than the full-spectrum dispersion relation.

An additional consideration should be made when comparing experimental data with (\ref{wavelengths}). While (\ref{wavelengths}) is the \textit{fastest}-growing wavelength, it is not the \textit{only} growing wavelength, even in the linear regime of pattern formation. Constant-dose irradiation experiments may lead to some coarsening of ripples before they can be observed \cite{castro-etal-PRB-2012}, thereby producing anomalously high wavelengths, if the total dose is not kept sufficiently low. It is therefore possible that some or all of the available wavelength data was measured outside of the proper linear growth regime \cite{castro-etal-PRB-2012}, leading to differences between observed ripple wavelengths at different fluences. Indeed, in \cite{castro-etal-PRB-2012}, the difference in $\lambda(\theta)$ for fixed $\theta$ is roughly a factor of 2 between fluences $6\times 10^{17} \frac{\text{ions}}{\text{cm}^2}$ and $6\times 10^{16} \frac{\text{ions}}{\text{cm}^2}$. In principle, this poses an obstacle to the comparison of theory and experiment. Throughout our collected wavelength data, we record the wavelengths at the lowest fluence possible.

With these caveats, we proceed with the expectation that the approximation (\ref{wavelengths}) is a reasonable, if imperfect, point of comparison with the experimental wavelength data--- which is itself \textit{also} possibly imperfect. 

\paragraph{Summary of results.} 

Our results are summarized in Tables \ref{table:results-wavelengths} and \ref{table:results-wavelengths-byauthor}. They show that the Ishii scaling (\ref{scaling-Ishii}) of $\eta^{-1}$ reduces $\chi^2_{\text{red}}$ by 13.7\% for Xe$^+ \to$ Si; by 18.5\% for Kr$^+ \to$ Ge; and by 42.5\% for Xe$^+ \to$ Ge. For Kr$^+ \to$ Si, where there is little experimental data, there is a negligible \textit{increase} by 0.012\%. 

The situation for Ar$^+ \to$ Si is more nuanced. When fitting \textit{all} wavelength data for this system, there is a roughly 10\% increase in $\chi^2_{\text{red}}$ when changing to the scaling of $\eta^{-1}$ implied by \cite{ishii-etal-JMR-2014}. However, we note that the experimental setup used in \cite{madi-etal-2008-PRL} is the same as in \cite{madi-thesis-2011}, where irradiation-induced heating up to 180 $^{\circ}$C is recorded; see Section 6.2 of \cite{madi-thesis-2011}. To the extent that we expect $\eta^{-1}$ to be determined by recombination of defects \cite{ishii-etal-JMR-2014}, we would \textit{also} expect that the recombination rate would be sensitive to temperature changes \cite{volkert-JAP-1991}. We therefore perform a second fit to the Ar$^+ \to$ Si data without the high-temperature data of \cite{madi-etal-2008-PRL}; an approximately 4\% improvement in fitting is then observed when using the form of the fluidity based on defect kinetics.

Additional insight is possible by breaking out the data by individual authors. For the wavelength measurements by \cite{madi-etal-2008-PRL,castro-etal-PRB-2012,hofsass-bobes-zhang-JAP-2016}, notable improvements in fitting are possible by switching from the scaling due to VM to that of Ishii. In particular, when taken on its own, the high-temperature data due to \cite{madi-etal-2008-PRL} is \textit{still} fit better with Ishii, suggesting that the same functional form applies even for higher temperatures, while defect recombination is likely affected, leading to a substantially different value of $P_{30}$. It also appears that the data due to \cite{ishii-thesis-2013} is an outlier, being much better fit with VM than Ishii. However, some of the fluxes used in \cite{ishii-thesis-2013} approach those for which \cite{madi-thesis-2011} recorded significant heating at lower energies. Excluding the data from \cite{ishii-thesis-2013} also, we find a roughly 14\% improvement in fitting the wavelength data using Ishii compared with VM, again consistent with the hypothesis of thermal effects. The data due to \cite{moreno-barrado-etal-PRB-2015} is also better fit with VM, possibly due to a range of fluxes used in the experiments, and appears to be the only true outlier that cannot be accounted for by thermal effects.

We conclude that the use of form (\ref{fluidity-ishii}) leads to broad improvements in the fits to the data, despite the caveats discussed above. This is a rather striking departure from the status quo, where $\eta^{-1}(\theta)$ has been assumed linear in $f$, implying $\lambda(\theta,E)$ flux-invariant. On the contrary, these results show that the flux, and \textit{not only the total fluence}, is an important quantity governing surface evolution. 
Finally, we note that it is possible, in principle, to see improvements in fitting to $\lambda(\theta)$ while using a nonphysical model. In the next section, we will use comparison with the steady stresses as a final model validation step.

\begin{table}[htpb]
\centering
\begin{tabular}{||c c c c c c c||} 
 \hline
Systems & VM, $\chi_{\text{red}}^2$ & Ishii, $\chi_{\text{red}}^2$ & change &  data, $\lambda(\theta)$ & $P_{30}$, VM & $P_{30}$, Ishii \\ [0.5ex]
 \hline\hline
 Ar$^+ \to$ Si (all) & 13.47 & 14.81 & +9.94\% & 27 & 2.1502 & 0.7589 \\
 Ar$^+ \to$ Si (no \cite{madi-etal-2008-PRL}) & 10.9798 & 10.5592 & -3.83\% & 22 & 1.9992 & 0.7238\\
 Ar$^+ \to$ Si (no \cite{madi-etal-2008-PRL,ishii-thesis-2013}) & 10.9164 & 9.3699 & -14.16\% & 18 & 1.8772 & 0.6847\\
 Kr$^+ \to$ Si & 294.0961 & 294.1311 & +0.012\% & 4 & 0.2099 & 0.0421\\
 Xe$^+ \to$ Si & 9.86 & 8.51 & -13.7\% & 9 & 0.7894 & 0.3064\\
 Kr$^+ \to$ Ge & 21.26 & 17.34 & -18.5\% & 6 & 6.2272 & 8.0291 \\
 Xe$^+ \to$ Ge & 0.0221 & 0.0127 & -42.5\% & 2 & 1.5265 & 1.4591 \\ [1ex] 
 \hline
\end{tabular}
\caption{Summary of fitting results for $\lambda(\theta)$ across energies, projectiles and targets. For some systems, fluxes are given as a range; in such cases, we simply use the midpoint.}
\label{table:results-wavelengths}
\end{table}

\begin{table}[htpb]
\centering
\begin{tabular}{||c c c c c ||} 
 \hline
Systems & VM, $\chi_{\text{red}}^2$ & Ishii, $\chi_{\text{red}}^2$ & change &  data, $\lambda(\theta)$ \\ [0.5ex]
 \hline\hline
Madi et al. \cite{madi-etal-2008-PRL} & 0.3738 & 0.2486 & -33.49\% & 5 \\
Castro et al. \cite{castro-etal-PRB-2012} & 6.0871 & 4.8449 & -20.41\% & 10 \\
Ishii \cite{ishii-thesis-2013} & 0.2589 & 2.1603 & +734.41\% & 4 \\
Moreno-Barrado et al. \cite{moreno-barrado-etal-PRB-2015} & 0.4525 & 0.6242 & +37.39\% & 5 \\ 
Hofs\"{a}ss et al. \cite{hofsass-bobes-zhang-JAP-2016} & 5.600 & 5.353 & -4.41\% & 3 \\ [1ex]
 \hline
\end{tabular}
\caption{Fitting results for $\lambda(\theta)$ for Ar$^+ \to$ Si with results broken out by author.}
\label{table:results-wavelengths-byauthor}
\end{table}

\subsection{Comparison to steady stresses}
\label{subsection:stresses}
Having fit $P_{10}, P_{20}, P_{30}, c_1, c_2$ and $c_3$, all \textit{dimensionless} quantities are fully-determined. We use the fit parameters from Ar$^+ \to$ Si where the high-temperature data \cite{madi-etal-2008-PRL} is excluded, while retaining all other data--- even the possible outliers, \cite{ishii-thesis-2013} and \cite{moreno-barrado-etal-PRB-2015}. 

In order to obtain \textit{dimensional} quantities, we will use $\frac{\rho_a}{\rho_c}$ as a fit parameter for the comparison between theoretical and experimental stresses. Physically, we require $\frac{\rho_a}{\rho_c} \in (0,1]$, as amorphization is widely understood to reduce density. Previous estimates place $\frac{\rho_a}{\rho_c}$ around 0.95 \cite{beardmore-gronbech-jensen-PRB-1999,hofsass-bobes-zhang-JAP-2016}. However, in order to give the fitter the freedom to choose nonphysical values, we allow $\frac{\rho_a}{\rho_c} \in (0,20]$, with values from $(1,20]$ expected to be nonphysical. In doing so, we gain confidence that finding physically-reasonable values of $\frac{\rho_a}{\rho_c}$ is due to the soundness of the underlying model, and not the tightness of our constraints. 

In Tables \ref{table:stresses-Ar-Si} and \ref{table:stresses-Ar-Si-offnormal}, we collect stress data for Ar$^+ \to$ Si across a variety of fluxes, three energies (250eV, 600eV, and 900eV) and two angles (normal incidence and 68$^{\circ}$). However, as before, we will exclude the 250eV data from this fit due to the significant heating during the experiments \cite{madi-thesis-2011}, which we anticipate will strongly influence $\eta^{-1}$ \cite{volkert-JAP-1991} by increasing the defect recombination rate $D_2$. For the 600eV and 900eV data, taken from \cite{ishii-thesis-2013,ishii-etal-JMR-2014}, it is reported that heating was controlled to within about 20 $^{\circ}$C of room temperature, at least for fluxes below $6.25 \frac{\text{ions}}{\text{nm}^2\cdot \text{s}}$.

We have the theoretical in-plane steady stress prediction for a fixed $E$ and $f$ and specified irradiation angle $\theta$ \cite{evans-norris-JPCM-2023,evans-norris-JEM-2024},
\begin{equation}
    \textbf{T}_{0xx}(\theta,E,f) = 6fA_D(E)\eta(\theta,E,f) + 2fA_I(E)\eta(\theta,E,f),
\end{equation}
where the continuum parameters are retrieved from the dimensionless fits as
\begin{equation}
\begin{gathered}
    A_D = \frac{\Omega Y(\theta)}{P_2h_0(\theta)}\bigg(\frac{\rho_c}{\rho_a}-1 \bigg) \\
    A_I = P_1A_D\bigg(\frac{\rho_a}{\rho_c}\bigg) \\
    \eta = \frac{\gamma \frac{\rho_c}{\rho_a}}{3P_3 fA_Dh_0(\theta)}.
\end{gathered}
\end{equation}

As in Section \ref{subsection:wavelengths}, we consider two possibilities for the scaling of the viscosity $\eta(\theta,E,f)$: one due to the VM form (\ref{fluidity-VM}), and the other due to the Ishii form (\ref{fluidity-ishii}). 
To fully determine $\frac{\rho_a}{\rho_c}$, we require an estimate for the surface energy $\gamma$ of amorphous Si. Elsewhere \cite{vauth-mayr-PRB-2007,vauth-mayr-PRB-2008b}, the value $\gamma=1.36 \frac{J}{\text{m}^2}$ has been computed, and leads to reasonable agreement with some experimental results \cite{norris-etal-NCOMM-2011,norris-PRB-2012-linear-viscous}. Another value, $\gamma=1.05 \pm 0.14 \frac{J}{\text{m}^2}$, exists within the literature \cite{hara-et-al-SS-2005} and has also been widely used. In what follows, we perform a fit to the Ar$^+ \to$ Si stress data using each of these two values with $\frac{\rho_a}{\rho_c}$ as the one remaining fit parameter. We also note that the inability to fit the stress data using the model developed thus far and a physically-reasonable value of $\frac{\rho_a}{\rho_c}$ would \textit{completely invalidate} the model.

\paragraph{Summary of results.} In Tables \ref{table:results-stresses-a}-\ref{table:results-stresses-c}, we summarize our results for the stresses using both the Ishii and VM models for $\eta^{-1}$. As with the wavelengths (Section \ref{subsection:wavelengths}), we find that the Ishii model of fluidity leads to significant improvements over VM in fitting to the stresses. The difference is even more pronounced when considering the stress data. Switching from VM to Ishii leads to a 70\%-80\% reduction in fitting error depending on choice of $\gamma$. At the same time, the Ishii model for $\eta^{-1}$ produces values of $\frac{\rho_a}{\rho_c}$ that are physically reasonable and closer to the expected value of around $0.95$ than the VM model. Moreover, taking the average of the two values given by \cite{vauth-mayr-PRB-2007,vauth-mayr-PRB-2008b} and \cite{hara-et-al-SS-2005} leads to a fit of $\frac{\rho_a}{\rho_c} \approx 0.96 \pm 0.03$, which aligns extremely well with our expectations and the measurements due to \cite{hofsass-bobes-zhang-JAP-2016}.

These results also show that apparent improvements in fitting $\lambda(\theta)$ by choosing the VM model of $\eta^{-1}$ are less consistent with a physical model.

\begin{table}[htpb]
\centering
\begin{tabular}{||c c c||} 
 \hline
$\eta^{-1}$ model & $\frac{\rho_a}{\rho_c}$ & $\chi^2_{red}$ \\ [0.5ex]
 \hline\hline
 VM & 0.917 $\pm$ 0.077 & 2982.78 \\
 Ishii & 1.011 $\pm$ 0.0361 & 842.69 \\ [1ex] 
 \hline
\end{tabular}
\caption{Results when using $\gamma=1.36 \frac{J}{\text{m}^2}$, the value due to \cite{vauth-mayr-PRB-2007,vauth-mayr-PRB-2008b}. Summary of fitting results for the in-plane stresses using both (\ref{fluidity-VM}) and (\ref{fluidity-ishii}).}
\label{table:results-stresses-a}
\end{table}

\begin{table}[htpb]
\centering
\begin{tabular}{||c c c||} 
 \hline
$\eta^{-1}$ model & $\frac{\rho_a}{\rho_c}$ & $\chi^2_{red}$ \\ [0.5ex]
 \hline\hline
 VM & 0.860 $\pm$ 0.078 & 2887.55 \\
 Ishii & 0.957 $\pm$ 0.0314 & 717.99 \\ [1ex] 
 \hline
\end{tabular}
\caption{Results when using $\gamma=1.225 \frac{J}{\text{m}^2}$, the midpoint of the values due to \cite{vauth-mayr-PRB-2007,vauth-mayr-PRB-2008b} and \cite{hara-et-al-SS-2005}. Summary of fitting results for the in-plane stresses using both (\ref{fluidity-VM}) and (\ref{fluidity-ishii}).}
\label{table:results-stresses-b}
\end{table}

\begin{table}[htpb]
\centering
\begin{tabular}{||c c c||} 
 \hline
$\eta^{-1}$ model & $\frac{\rho_a}{\rho_c}$ & $\chi^2_{red}$ \\ [0.5ex]
 \hline\hline
 VM & 0.784 $\pm$ 0.0622 & 2768.57 \\
 Ishii & 0.885 $\pm$ 0.0265 & 602.61 \\ [1ex] 
 \hline
\end{tabular}
\caption{Results when using $\gamma=1.05 \frac{J}{\text{m}^2}$, the value due to \cite{hara-et-al-SS-2005}. Summary of fitting results for the in-plane stresses using both (\ref{fluidity-VM}) and (\ref{fluidity-ishii}).}
\label{table:results-stresses-c}
\end{table}

\section{Discussion}
\label{section:discussion}


\paragraph{Conclusions.} We have developed a continuum model of an irradiated Si or Ge surface as a viscous thin film. The effect of ion implantation is incorporated by two modifications to the bulk conservation laws: first, through an anisotropic component of the stress tensor, and, second, as an equation of state that models loss of density due to ion implantation. The interfaces are modeled according to \cite{evans-norris-JEM-2024} and boundary conditions appropriate for moving interfaces undergoing phase change are applied.

The linear dispersion relation resulting from analysis of this model implies a critical angle $\theta_c$ of irradiation, beyond which surface perturbations are expected to grow, and ripples become visible in experiments. We have surveyed the existing experimental literature for $\theta_c$ determined for Ar$^+$, Kr$^+$ and Xe$^+$ irradiation of Si and Ge targets from 250eV to 2000eV, and used this data to perform fits to our growth rate model. Fit parameters were constrained in order to respect experimental trends noted elsewhere in the literature. From these fits, reasonable parameter values were extracted.

Then, appending the quartic term for viscous surface relaxation \cite{orchard-ASR-1962} to our growth model, we fit the resulting theoretical values of $\lambda(\theta)$ given two assumptions about the ion-enhanced fluidity $\eta^{-1}(\theta)$: one assuming a linear dependence on the flux $f$, and the other assuming a square-root dependence on $f$. Despite having received very little attention, the model of \cite{ishii-etal-JMR-2014} for $\eta^{-1}(\theta)$ leads to significantly improved fits to the existing $\lambda(\theta)$ data. Moreover, $\eta^{-1}(\theta) \propto \sqrt{f}$ enables a theoretical explanation for the apparent flux-dependence of the in-plane stresses at normal incidence--- an experimental observation that cannot be explained using the widely-used linear scaling in $f$.

Collectively, these results lend credibility to our parameter values and model (\ref{dispnrelnwBA}), which are shown to be broadly consistent with the experimental literature on three commonly-measured quantities: $\theta_c$, $\lambda(\theta)$ and the in-plane stress. We also find that modeling $\eta^{-1}(\theta)$ based on creation and recombination of Frenkel pairs \cite{ishii-etal-JMR-2014} makes this parsimony possible, and the \textit{status quo} treatment of $\eta^{-1}(\theta)$ linear in $f$ and $E$ should evidently be abandoned.

These results also illustrate the value of the present approach. While there are many experimental parameters that can be adjusted, experimental results can be used alongside theoretical models to discern active mechanisms--- such as the defect dynamics of \cite{ishii-etal-JMR-2014} which motivate the form (\ref{scaling-Ishii}). With enough data and a physically-grounded model, it is possible, in principle, to begin to estimate otherwise-unmeasurable quantities, like $\eta^{-1}(0)$.

Finally, it is also striking that we have obtained the present results while completely excluding the effects of erosion and redistribution, widely regarded as main drivers of ion-induced pattern formation. For the first time, we have shown that a viscous flow model \textit{on its own} is capable of parsimoniously explaining a large swath of experimental data for noble gas irradiation of Si and Ge at low energies.

\paragraph{Open questions and future work.}

\begin{itemize}
    \item Given the evident importance of correctly accounting for defect kinetics, We propose that the model of \cite{ishii-etal-JMR-2014} for $\eta^{-1}(\theta)$ should be studied and extended. Our analysis also suggests that thermal effects due to radiation-induced heating of the target are likely to be important. In particular, we find that the model for fluidity based on defect kinetics leads to a superior fit for the high-temperature wavelength data \cite{madi-etal-2008-PRL} when it is isolated from the lower-temperature wavelength data. This finding is consistent with discussion elsewhere \cite{volkert-JAP-1991}.
    \item In the present work, APF and IIS have been incorporated on a phenomenological basis, with no particular mechanism in mind, although it is plausible \cite{evans-norris-JPCM-2023} that IIS is due to local changes in density induced by the accumulation of defects. Development of a fully mechanistic description of APF and IIS would allow the extension of the present model to cases where patterns are known not to form: Ne$^+$ on Si, and Ne$^+$, Ar$^+$ on Ge. Our results show that APF and IIS should exhibit the same scaling with energy, suggesting common cause.
    \item We have considered only the energy range from 250eV-2000eV for experimental systems where reliable $\theta_c$ data exists. Although the model (\ref{dispnrelnwBA}) leads to good agreement with the experimental systems considered here, it is possible that the agreement we observe breaks down outside of this energy range. Filling in the experimental literature with $\theta_c$, $\lambda(\theta)$ and $\textbf{T}_{0xx}$ data for systems not considered in the present work would be highly valuable.
    \item The present work accounts for the strong suppression of ripple formation for Ar$^+ \to$ Si near 1500eV. However, we have made no effort to account for the apparent \textit{return} of ripple patterns for Ar$^+ \to$ Si at energies greater than around 20keV \cite{hofsass-bobes-zhang-JAP-2016}. This is for three main reasons. 
    \begin{itemize}
        \item There is a lack of experimental data on $\theta_c$ and $\lambda(\theta)$ in the intermediate energy range of 2keV-20keV for the projectile-target combinations considered here, which would otherwise enable comparison between experiment and theory. 
        \item It is also possible that the underlying dependence of certain parameter groups on $E$ is non-monotonic for as-yet-unknown physical reasons. As-is, the energy dependence of $A_D$ and $A_I$ resulting from our fits would appear to imply that $A_D$ and $A_I$ become negligibly small at the energy increases toward 20keV. Pattern formation would then appear to require either non-monotonicity for $A_I$ and $A_D$, or other destabilizing mechanisms to dominate, such as erosion and redistribution.
        \item While we have assumed that $\frac{\rho_a}{\rho_c}$ is a constant, this assumption may break down at significantly higher energies \cite{beardmore-gronbech-jensen-PRB-1999}. If $\frac{\rho_a}{\rho_c}$ approaches 1 after a sufficiently high energy is reached, Boundary Amorphization would be expected to significantly weaken. We propose that this could happen if increased heating leads to much faster annealing of defects, hence recrystallization and re-densification of the target. These effects, combined with the cross-beam standard deviation of ion implantation $\beta$ approaching 0, could sharply decrease $\theta_c$ \cite{evans-norris-JPCM-2023}. 
    \end{itemize}
    \item In the present work, erosion is incorporated only as an interfacial velocity which removes mass from the free surface and induces phase change at the amorphous-crystalline interface--- hence no instability can be caused by these prompt-regime mechanisms in this model. That such a model can evidently produce such a parsimonious explanation of a large swath of experimental data is an \textit{extremely surprising} departure from the prevailing explanations for ion-induced nanopatterning. Indeed, even \cite{norris-etal-SREP-2017}, where it was shown that erosive and viscous effects could occur at comparable magnitudes for 1keV Ar$^+ \to$ Si, was a surprise when first published. The present work shows that prompt-regime effects are fully \textit{unnecessary} for the experimental systems considered here, except for describing the Boundary Amorphization effect.
    \item Computed values of coefficients within the Crater Function Framework \cite{norris-etal-NCOMM-2011,norris-arXiv-2014-pycraters,norris-etal-SREP-2017} are large enough that the linear superposition of those terms with the viscous flow terms of the present model \textit{should} lead to the viscous flow terms being entirely ``drowned out". This raises questions as to the coupling of atomistic processes occurring on the fast time scale (erosion, redistribution), and continuum processes occurring on the slow time scale (viscous flow, defect kinetics). Addressing this matter will be a topic of forthcoming work.
\end{itemize}

\section*{Acknowledgments}
TPE wishes to thank the National Science Foundation for its generous support of this work through DMS-1840260 while at Southern Methodist University and DMS-2136198 while at University of Utah.

\clearpage

\appendix

\section{Figures: comparison of model and data}
\begin{figure}[htpb]
	\centering	
 \includegraphics[totalheight=6cm]{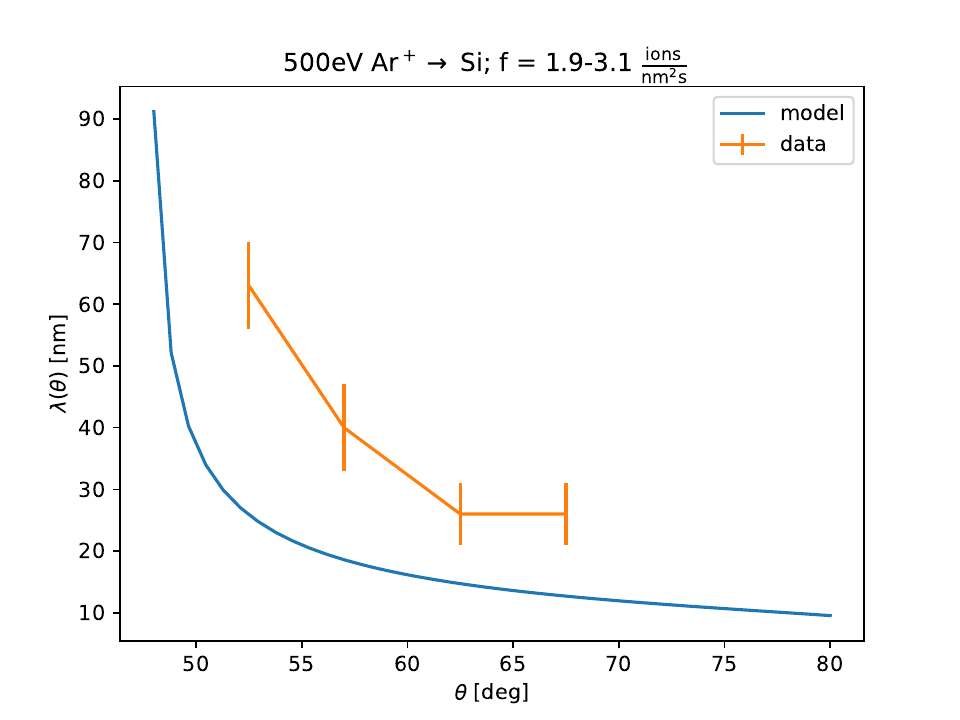}
 \includegraphics[totalheight=6cm]{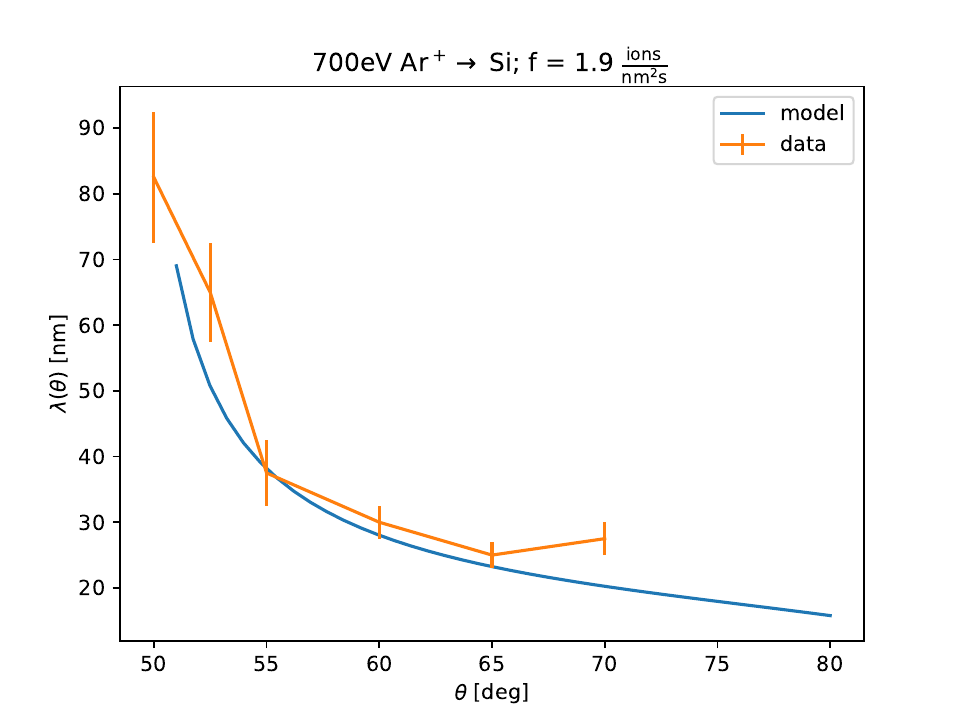}
\includegraphics[totalheight=6cm]{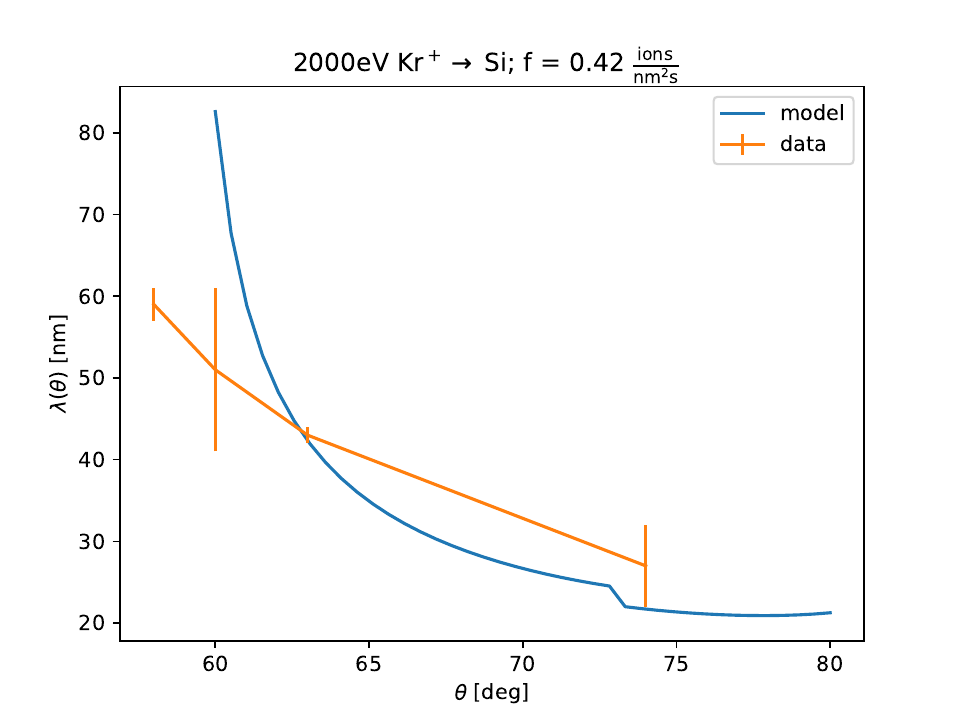}
\includegraphics[totalheight=6cm]{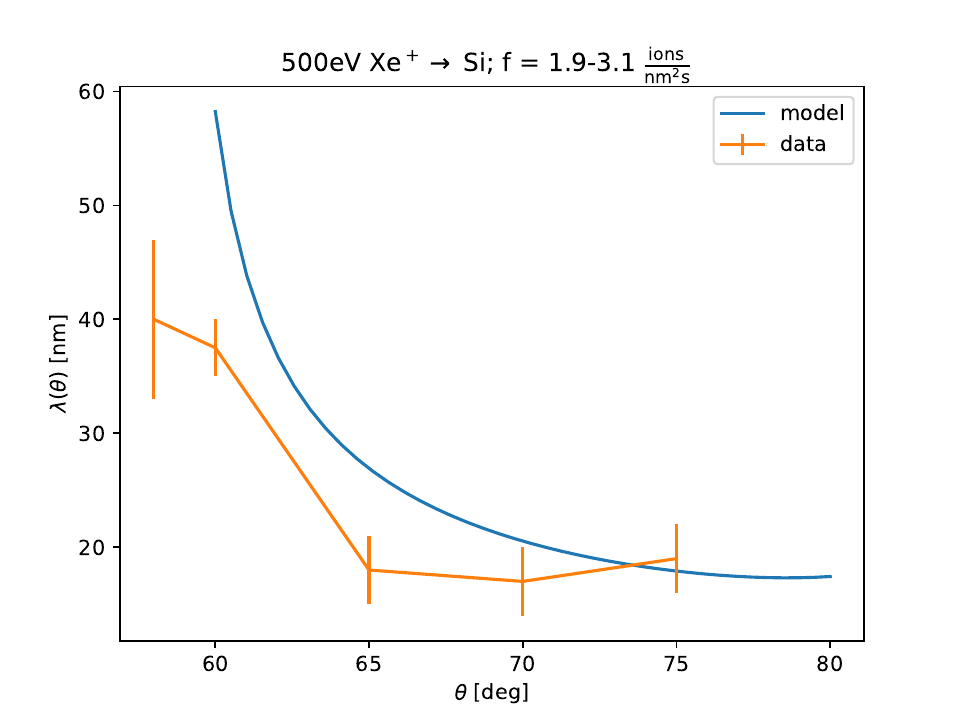}
\caption{Model vs. wavelength data for Si targets. The fits use $P_1$ and $P_2$ from fitting to $\theta_c$. $P_3$ assumes Ishii's model of the ion-enhanced fluidity.} \label{fig:lambda-results-Si}
\end{figure}

\begin{figure}[htpb]
	\centering	
\includegraphics[totalheight=6cm]{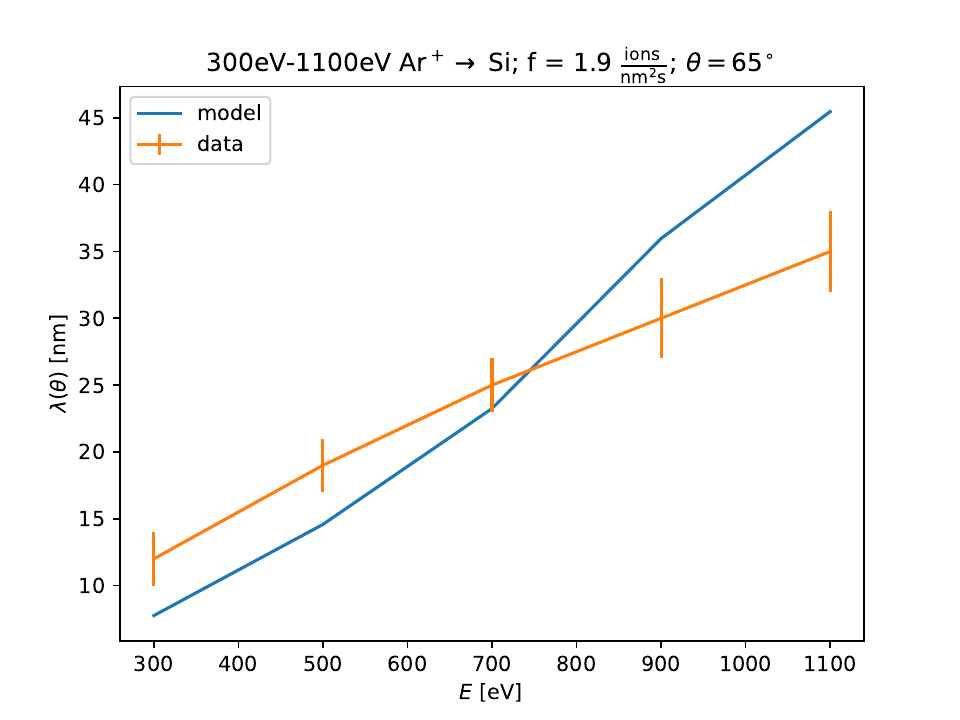}
\caption{Model vs wavelength data for Ar$^+ \to$ Si at 65 degrees and various energies. The fits use $P_1$ and $P_2$ from fitting to $\theta_c$. $P_3$ assumes Ishii's model of the ion-enhanced fluidity. Data is from \cite{castro-etal-PRB-2012}. } \label{fig:lambda-results-65degrees}
\end{figure}

\begin{figure}[htpb]
	\centering	
\includegraphics[totalheight=6cm]{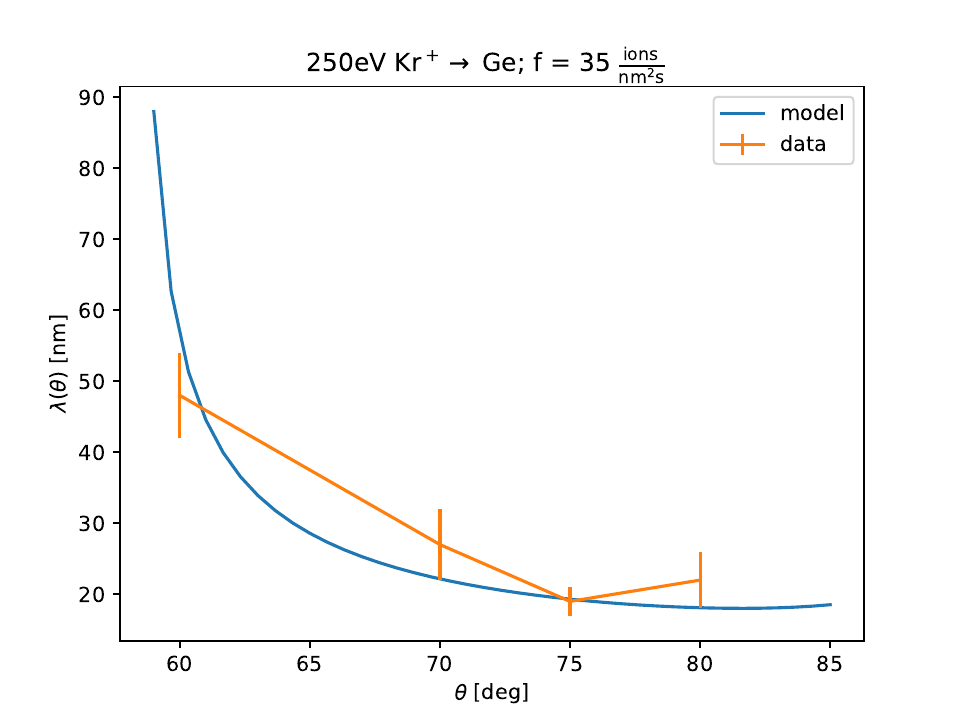}
\includegraphics[totalheight=6cm]{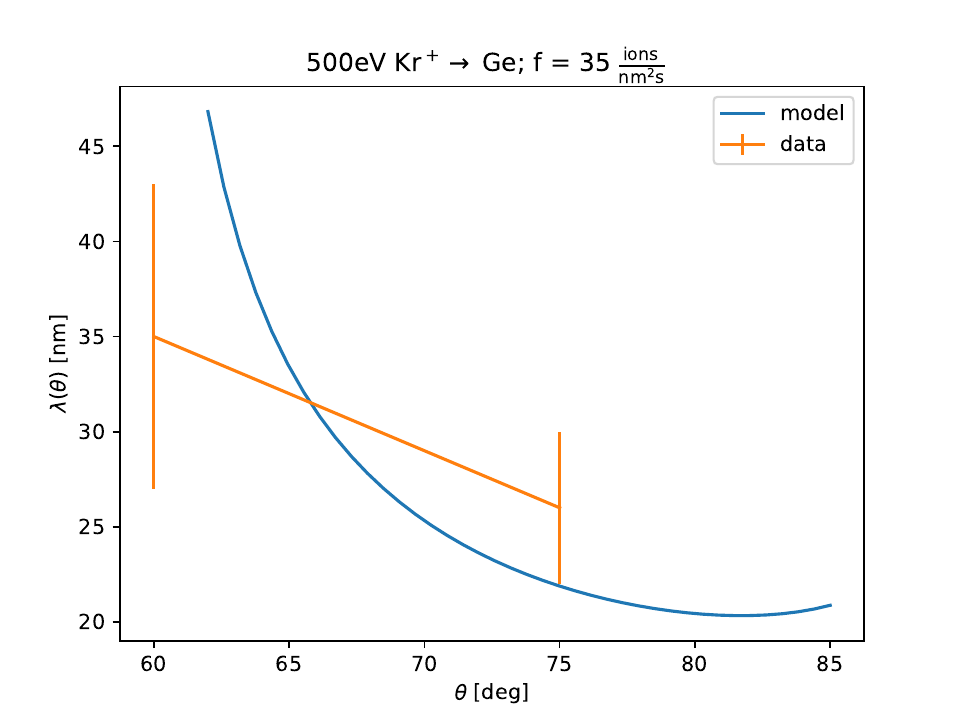}
\includegraphics[totalheight=6cm]{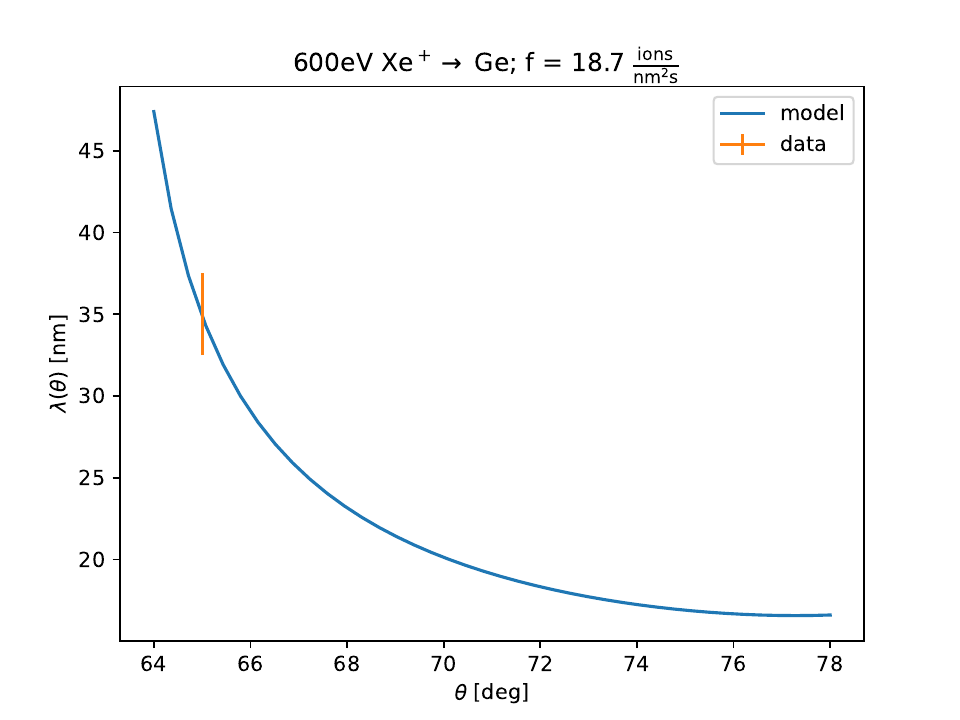}
\includegraphics[totalheight=6cm]{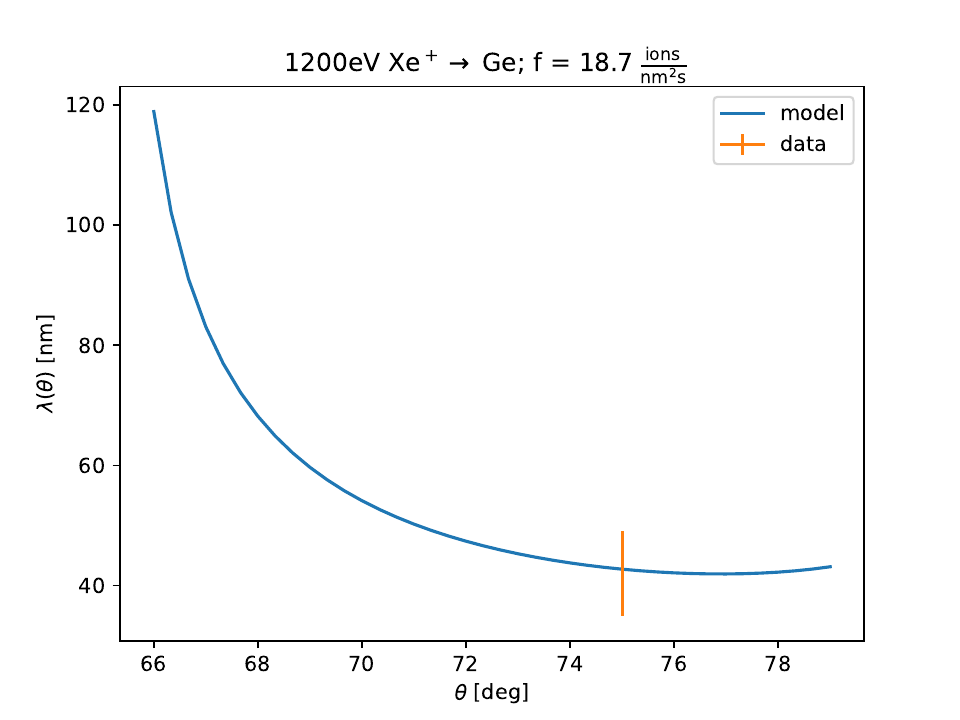}
\caption{Model vs. wavelength data for Ge targets. The fits use $P_1$ and $P_2$ from fitting to $\theta_c$. $P_3$ assumes Ishii's model of the ion-enhanced fluidity.  } \label{fig:lambda-results-Ge}
\end{figure}

\begin{figure}[htpb]
	\centering	
\includegraphics[totalheight=6cm]{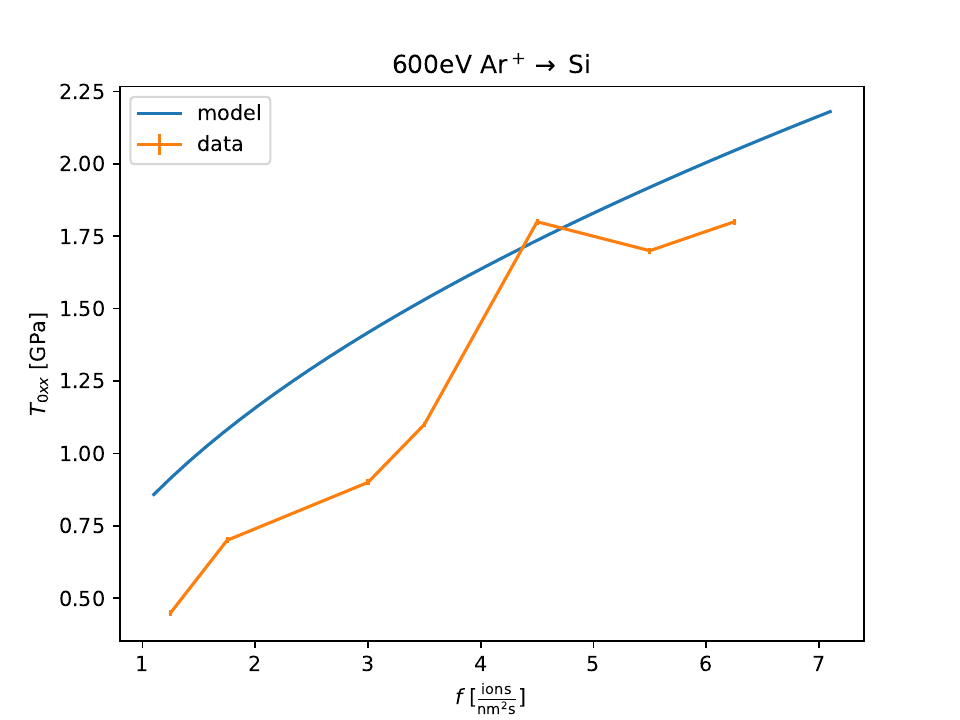}
\includegraphics[totalheight=6cm]{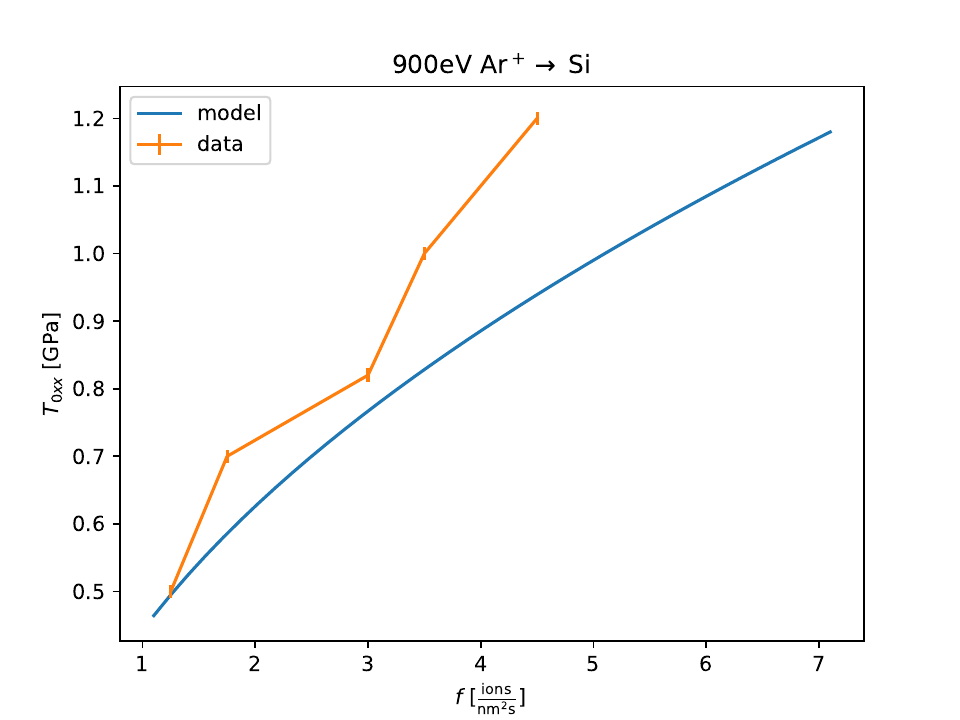}
\caption{Comparison of model and stress data for normal-incidence Ar$^+ \to$ Si. Here, we assume $\gamma = 1.36 \frac{J}{\text{m}^2}$ and use the fit parameters described in Section \ref{subsection:stresses}. Data is from \cite{ishii-thesis-2013,ishii-etal-JMR-2014}. } \label{fig:stress-results}
\end{figure}

\clearpage

\section{Collected data: critical angles, wavelengths, and in-plane stress}
In the main text, we fit $\theta_c$ for each of the systems in Tables \ref{table:Si-data} and \ref{table:Ge-data} using (\ref{dispnrelnwBA}) as a theoretical model.  When experimental data is uncertain, we use the uncertainty range specified in the source reference. If no uncertainty range is given--- for example, because determining $\theta_c$ was not the purpose of the work cited, as in \cite{ziberi-etal-PRB-2005}--- then we assume that $\theta_c$ is simply the average of the two angles $\theta$ between which patterning has begun. For consistency, this approach is taken even when visual inspection suggests that it might worsen the quality of the fit.

We summarize the experimental literature on $\lambda(\theta)$ for pattern-forming systems among Ar$^+$, Kr$^+$, Xe$^+$ on Si and Ge for 250eV-2000eV in Table \ref{table:wavelengths-Si-Ar}-\ref{table:wavelengths-Ge-Xe}. In Section \ref{section:results}, these values are used for comparison with the final, fitted model.

In Tables \ref{table:stresses-Ar-Si} and \ref{table:stresses-Ar-Si-offnormal}, we collect experimental measurements of in-plane stress for Ar$^+ \to$ Si at various energies and fluxes and at two angles of incidence: normal, and 68$^{\circ}$ off-normal. This data is useful in our discussion of the ion-enhanced fluidity $\eta^{-1}(\theta)$ and theoretical form of the in-plane stress $\textbf{T}_{0xx}$. We note that the sources for this data \cite{madi-thesis-2011,ishii-thesis-2013,ishii-etal-JMR-2014} do not report error bars on their stress measurements. For the purposes of fitting, we therefore assume that the data is \textit{exact}, and apply only $\pm .01$ GPa to each data point as a trivial amount of uncertainty.

In all tables, $\theta$ is reported in degrees away from normal incidence. Flux $f$ is reported as the nominal flux through a plane perpendicular to the ion beam (as opposed to an ``effective flux", which is sometimes reported). 

\begin{table}[htpb]
\parbox{.45\linewidth}{
\centering
\begin{tabular}{||c c c c c ||}
 \hline
 Target & Ion & E(eV) & $\theta_c$ & Source \\ [0.5ex] 
 \hline\hline
 Si & Ar & 250 & 46-50 & \cite{madi-etal-JPCM-2009} \\  
 \hline
 Si & Ar & 300 & 45-50 & \cite{castro-etal-PRB-2012} \\
 \hline
 Si & Ar & 500 & 44-48 & \cite{moreno-barrado-etal-PRB-2015} \\
 \hline
 Si & Ar & 650 & 30-65 & \cite{hofsass-bobes-zhang-JAP-2016} \\
 \hline
 Si & Ar & 700 & 47-53 & \cite{castro-etal-PRB-2012} \\
 \hline
 Si & Ar & 1000 & 45-50 & \cite{madi-etal-PRL-2011} \\
 \hline
 Si & Ar & 1300 & 60-65 & \cite{hofsass-bobes-zhang-JAP-2016} \\
 \hline
 Si & Ar & 1500 & 60-75 & \cite{ziberi-etal-PRB-2005} \\
 \hline
 Si & Kr & 1000 & 56-60 & \cite{perkinsonthesis2017} \\
 \hline
 Si & Kr & 1200 & 60-65 & \cite{frost-etal-APA-2008} \\
 \hline
 Si & Kr & 2000 & 55-60 & \cite{ziberi-etal-PRB-2005,engler-etal-PRB-2014,seo-et-al-JPCM-2022} \\
 \hline
Si & Xe & 300 & 30-65 & \cite{moreno-barrado-etal-PRB-2015} \\
 \hline
Si & Xe & 500 & 56-60 & \cite{moreno-barrado-etal-PRB-2015} \\
 \hline
Si & Xe & 700 & 56-60 & \cite{moreno-barrado-etal-PRB-2015} \\
 \hline
 Si & Xe & 1000 & 56-60 & \cite{moreno-barrado-etal-PRB-2015} \\
 \hline
 Si & Xe & 2000 & 60-70 & \cite{ziberi-etal-PRB-2005} \\
 \hline
\end{tabular}
\caption{$\theta_c$ data for Ar$^+$, Kr$^+$, Xe$^+$ $\to$ Si from 250eV to 2000eV}\label{table:Si-data}
}  
\hfill
\parbox{.45\linewidth}{
\centering
\begin{tabular}{||c c c c c ||}
 \hline
 Target & Ion & E (eV) & $\theta_c$ & Source \\ [0.5ex] 
 \hline\hline
 Ge & Kr & 250 & 56-60 & \cite{perkinson-JVSTA-2013-Kr-Ge} \\  
 \hline
 Ge & Kr & 500 & 56-60 & \cite{perkinson-JVSTA-2013-Kr-Ge}  \\
 \hline
 Ge & Kr & 1000 & 60-65 & \cite{anzenberg-etal-PRB-2012} \\
 \hline
 Ge & Kr & 1200 & 60-70 & \cite{Teichmann2013} \\
 \hline
 Ge & Kr & 2000 & 60-75 & \cite{ziberi-etal-PRB-2005}  \\
 \hline
 Ge & Xe & 300 & 61-63 & \cite{hans-etal-NT-2022} \\
 \hline
 Ge & Xe & 400 & 61-65 & \cite{Teichmann2013, hans-etal-NT-2022} \\
 \hline
 Ge & Xe & 500 & 61-63 & \cite{hans-etal-NT-2022} \\
 \hline
 Ge & Xe & 600 & 61-65 & \cite{Teichmann2013,hans-etal-NT-2022} \\
 \hline
 Ge & Xe & 800 & 61-65 & \cite{Teichmann2013,hans-etal-NT-2022} \\
 \hline
Ge & Xe & 1200 & 65-70 & \cite{Teichmann2013} \\
 \hline
 Ge & Xe & 2000 & 65-75 & \cite{ziberi-etal-PRB-2005,Teichmann2013} \\
 \hline
\end{tabular}
\caption{$\theta_c$ data for Kr$^+$, Xe$^+$ $\to$ Ge from 250eV to 2000eV}  \label{table:Ge-data}

}

\end{table}

\begin{table}[htpb]
\centering
\begin{tabular}{||c c c c c c c ||} 
 \hline
Target & Ion & E (eV) & $\theta$ (deg) & f ($\frac{\text{ions}}{\text{nm}^2 \text{s}}$) & $\lambda(\theta)$ (nm) & Source \\ [0.5ex] 
 \hline\hline
Si  & Ar & 250 & 50 & 36  & $48 \pm 7$  & \cite{madi-etal-2008-PRL} \\ 
Si  & Ar & 250 & 55 & 36  & $35 \pm 5$ & \cite{madi-etal-2008-PRL}\\
Si  & Ar & 250 & 60 & 36  & $22 \pm 3$ & \cite{madi-etal-2008-PRL}\\
Si  & Ar & 250 & 65 & 36 & $20 \pm 3$  & \cite{madi-etal-2008-PRL}\\
Si  & Ar & 250 & 70 & 36 & $20 \pm 3$ & \cite{madi-etal-2008-PRL}\\ 
Si  & Ar & 300 & 65 & 1.9 & $12 \pm 2$ & \cite{castro-etal-PRB-2012}\\ 
Si  & Ar & 500 & 50 & 1.9-3.1 & $82 \pm 10$ & \cite{moreno-barrado-etal-PRB-2015} \\ 
Si  & Ar & 500 & 52.5 & 1.9-3.1 & $63 \pm 7$ & \cite{moreno-barrado-etal-PRB-2015} \\
Si  & Ar & 500 & 57 & 1.9-3.1 & $40 \pm 7$ & \cite{moreno-barrado-etal-PRB-2015}\\
Si  & Ar & 500 & 62.5 & 1.9-3.1 & $26 \pm 5$ & \cite{moreno-barrado-etal-PRB-2015}\\
Si  & Ar & 500 & 65 & 1.9 & $19 \pm 2$ & \cite{castro-etal-PRB-2012} \\
Si  & Ar & 500 & 67.5 & 1.9-3.1 & $26 \pm 5$ & \cite{moreno-barrado-etal-PRB-2015}\\
Si  & Ar & 600 & 68 & 1.25 & $39 \pm 7$ & \cite{ishii-thesis-2013}, Figure 4.7\\
Si  & Ar & 600 & 68 & 5 & $35 \pm 7$ & \cite{ishii-thesis-2013}, Figure 4.7\\
Si  & Ar & 600 & 68 & 10 & $31 \pm 7$ & \cite{ishii-thesis-2013}, Figure 4.7\\
Si  & Ar & 600 & 68 & 20 & $33 \pm 4$ & \cite{ishii-thesis-2013}, Figure 4.7\\
Si  & Ar & 650 & 65 & 1.9-3.8 & $31 \pm 10$ & \cite{hofsass-bobes-zhang-JAP-2016}\\
Si  & Ar & 700 & 50 & 1.9 & $82.5 \pm 10$ & \cite{castro-etal-PRB-2012} \\ 
Si  & Ar & 700 & 52.5 & 1.9 & $65 \pm 7.5$ & \cite{castro-etal-PRB-2012} \\
Si  & Ar & 700 & 55 & 1.9 & $37.5 \pm 5$ & \cite{castro-etal-PRB-2012} \\
Si  & Ar & 700 & 60 & 1.9 & $30 \pm 2.5$ &\cite{castro-etal-PRB-2012} \\
Si  & Ar & 700 & 65 & 1.9  & $25 \pm 2$ &\cite{castro-etal-PRB-2012} \\
Si  & Ar & 700 & 70 & 1.9 & $27.5 \pm 2.5$ &\cite{castro-etal-PRB-2012} \\
Si  & Ar & 900 & 65 & 1.9 & $30 \pm 3$ &\cite{castro-etal-PRB-2012} \\
Si  & Ar & 1000 & 65 & 1.9-3.8 & $40 \pm 10$ &\cite{hofsass-bobes-zhang-JAP-2016} \\
Si  & Ar & 1100 & 65 & 1.9 & $35 \pm 3$ &\cite{castro-etal-PRB-2012} \\
Si  & Ar & 1300 & 65 & 1.9-3.8 & $38 \pm 10$ &\cite{hofsass-bobes-zhang-JAP-2016} \\
[1ex] 
 \hline
\end{tabular}
\caption{ From \cite{castro-etal-PRB-2012}, we use the data associated with a lower fluence, which is therefore more likely in the proper linear regime.}
\label{table:wavelengths-Si-Ar}
\end{table}

\begin{table}[htpb]
\centering
\begin{tabular}{||c c c c c c c ||} 
 \hline
Target & Ion & E (eV) & $\theta$ (deg) & f ($\frac{\text{ions}}{\text{nm}^2 \text{s}}$) & $\lambda(\theta)$ (nm) & Source \\ [0.5ex] 
 \hline\hline
Si  & Kr & 2000 & 58 & 0.42  & $59 \pm 2$ & \cite{engler-etal-PRB-2014} \\ 
Si  & Kr & 2000 & 60 & 0.42  & $51 \pm 10$ & \cite{engler-etal-PRB-2014} \\ 
Si  & Kr & 2000 & 63 & 0.42  & $43 \pm 1$ & \cite{engler-etal-PRB-2014} \\ 
Si  & Kr & 2000 & 74 & 0.62  & $27 \pm 5$ & \cite{seo-et-al-JPCM-2022} \\ 
[1ex] 
 \hline
\end{tabular}
\caption{Kr$^+ \to$ Si wavelength data. }
\label{table:wavelengths-Si-Kr}
\end{table}

\begin{table}[htpb]
\centering
\begin{tabular}{||c c c c c c c ||} 
 \hline
Target & Ion & E (eV) & $\theta$ (deg) & f ($\frac{\text{ions}}{\text{nm}^2 \text{s}}$) & $\lambda(\theta)$ (nm) & Source \\ [0.5ex] 
 \hline\hline
Si  & Xe & 300 & 65 & 1.9-3.1  & $12.5 \pm 2.5$ & \cite{moreno-barrado-etal-PRB-2015} \\ 
Si  & Xe & 500 & 58 & 1.9-3.1  & $40 \pm 7$ & \cite{moreno-barrado-etal-PRB-2015} \\ 
Si  & Xe & 500 & 60 & 1.9-3.1  & $37.5 \pm 2.5$ & \cite{moreno-barrado-etal-PRB-2015} \\
Si  & Xe & 500 & 65 & 1.9-3.1  & $18 \pm 3$ & \cite{moreno-barrado-etal-PRB-2015}\\
Si  & Xe & 500 & 70 & 1.9-3.1  & $17 \pm 3$ & \cite{moreno-barrado-etal-PRB-2015}\\
Si  & Xe & 500 & 75 & 1.9-3.1  & $19 \pm 3$  & \cite{moreno-barrado-etal-PRB-2015}\\ 
Si  & Xe & 700 & 65 & 1.9-3.1  & $17.5 \pm 2.5$ & \cite{moreno-barrado-etal-PRB-2015} \\ 
Si  & Xe & 900 & 65 & 1.9-3.1  & $22.5 \pm 2.5$ & \cite{moreno-barrado-etal-PRB-2015} \\ 
Si  & Xe & 1000 & 65 & 1.9-3.1 & $25 \pm 2.5$ & \cite{moreno-barrado-etal-PRB-2015} \\ [1ex] 
 \hline
\end{tabular}
\caption{ Xe$^+ \to$ Si wavelength data.}
\label{table:wavelengths-Si-Xe}
\end{table}

\begin{table}[htpb]
\centering
\begin{tabular}{||c c c c c c c ||} 
 \hline
Target & Ion & E (eV) & $\theta$ (deg) & f ($\frac{\text{ions}}{\text{nm}^2 \text{s}}$)  & $\lambda(\theta)$ (nm) & Source \\ [0.5ex] 
 \hline\hline
Ge  & Kr & 250 & 60 & 35  & $48 \pm 6$  & \cite{perkinson-JVSTA-2013-Kr-Ge} \\ 
Ge  & Kr & 250 & 70 & 35 & $27 \pm 5$ & \cite{perkinson-JVSTA-2013-Kr-Ge}\\
Ge  & Kr & 250 & 75 & 35 &  $19 \pm 2$ & \cite{perkinson-JVSTA-2013-Kr-Ge}\\
Ge  & Kr & 250 & 80 & 35 & $22 \pm 4$ & \cite{perkinson-JVSTA-2013-Kr-Ge}\\
Ge  & Kr & 500 & 60 & 35 &  $35 \pm 8$  & \cite{perkinson-JVSTA-2013-Kr-Ge}\\ 
Ge  & Kr & 500 & 75 & 35 & $26 \pm 4$ & \cite{perkinson-JVSTA-2013-Kr-Ge} \\ 
[1ex] 
 \hline
\end{tabular}
\caption{Kr$^+ \to$ Ge wavelength data. For these experiments, the same chamber as in \cite{madi-etal-2008-PRL,madi-thesis-2011} was used. Based on this, \cite{perkinson-JVSTA-2013-Kr-Ge}, notes that a similar radiation-induced heating up to around 180 $^{\circ}$C is expected, although the temperature was evidently not directly measured.}
\label{table:wavelengths-Ge-Kr}
\end{table}

\begin{table}[htpb]
\centering
\begin{tabular}{||c c c c c c c ||} 
 \hline
Target & Ion & E (eV) & $\theta$ (deg) & f ($\frac{\text{ions}}{\text{nm}^2 \text{s}}$)  & $\lambda(\theta)$ (nm) & Source \\ [0.5ex] 
 \hline\hline
Ge  & Xe & 600 & 65 & 18.7 & $35 \pm 5$ & \cite{Teichmann2013} \\ 
Ge  & Xe & 1200 & 75 & 18.7 & $42 \pm 7$ & \cite{Teichmann2013} \\
[1ex] 
 \hline
\end{tabular}
\caption{Xe$^+ \to$ Ge wavelength data.  }
\label{table:wavelengths-Ge-Xe}
\end{table}

\begin{table}[htpb]
\centering
\begin{tabular}{||c c c c c c ||} 
 \hline
Target & Ion & E (eV) & f ($\frac{\text{ions}}{\text{nm}^2 \text{s}}$) & In-plane Stress (GPa) & Source \\ [0.5ex] 
 \hline\hline
Si  & Ar & 250 & 12 & 1.3 & \cite{madi-thesis-2011} \\ 
Si  & Ar & 250 & 35 & 1.6 & \cite{madi-thesis-2011} \\
Si  & Ar & 600 & 1.25 & 0.45 & \cite{ishii-etal-JMR-2014} \\
Si  & Ar & 600 & 1.75 & 0.7 & \cite{ishii-etal-JMR-2014}\\
Si  & Ar & 600 & 3  & 0.9 & \cite{ishii-etal-JMR-2014}\\ 
Si  & Ar & 600 & 3.5  & 1.1 & \cite{ishii-etal-JMR-2014}\\ 
Si  & Ar & 600 & 4.5  & 1.8 & \cite{ishii-etal-JMR-2014} \\
Si  & Ar & 600 & 5.5  & 1.7 & \cite{ishii-etal-JMR-2014} \\
Si  & Ar & 600 & 6.25  & 1.8 & \cite{ishii-etal-JMR-2014} \\
Si  & Ar & 900 & 1.25  & 0.5 & \cite{ishii-etal-JMR-2014} \\
Si  & Ar & 900 & 1.75  & 0.7 & \cite{ishii-etal-JMR-2014} \\
Si  & Ar & 900 & 3  & 0.8 & \cite{ishii-etal-JMR-2014} \\
Si  & Ar & 900 & 3.5  & 1.0 & \cite{ishii-etal-JMR-2014} \\
Si  & Ar & 900 & 4.5  & 1.2 & \cite{ishii-etal-JMR-2014} \\
[1ex] 
 \hline
\end{tabular}
\caption{In-plane stress due to Ar$^+ \to$ Si irradiation at various energies and fluxes. All stresses are recorded at normal incidence. In \cite{madi-thesis-2011}, heating to 200 $^{\circ}$C was applied; we therefore exclude this data from the fitting described in the main text due to the expectation that the defect recombination rate is temperature-sensitive. It is recorded here only for completeness, and for the role that it has played elsewhere in the literature \cite{norris-PRB-2012-linear-viscous}. }
\label{table:stresses-Ar-Si}
\end{table}

\begin{table}[htpb]
\centering
\begin{tabular}{||c c c c c c ||} 
 \hline
Target & Ion & E (eV) & f ($\frac{\text{ions}}{\text{nm}^2 \text{s}}$) & In-plane Stress (GPa) & Source \\ [0.5ex] 
 \hline\hline
Si  & Ar & 600 & 4  & 0.42 & \cite{ishii-thesis-2013} (Figure 4.3) \\
Si  & Ar & 600 & 5  & 0.375 & \cite{ishii-thesis-2013} (Figure 4.3) \\
Si  & Ar & 600 & 9  & 0.5 & \cite{ishii-thesis-2013} (Figure 4.3) \\
Si  & Ar & 600 & 10  & 0.55 & \cite{ishii-thesis-2013} (Figure 4.3) \\
Si  & Ar & 600 & 11.25  & 1 & \cite{ishii-thesis-2013} (Figure 4.3) \\
Si  & Ar & 600 & 12.5  & 1.1 & \cite{ishii-thesis-2013} (Figure 4.3) \\
[1ex] 
 \hline
\end{tabular}
\caption{In-plane stress due to Ar$^+ \to$ Si irradiation at various energies and fluxes. All stresses are recorded at 68$^{\circ}$.}
\label{table:stresses-Ar-Si-offnormal}
\end{table}

\clearpage

\bibliographystyle{plain}
\bibliography{bapaperrefs2}

\end{document}